\documentclass[12pt]{article}

\usepackage{epsfig}

\usepackage{amssymb, amsmath,bbold}
\usepackage{graphicx}
\usepackage{overpic}

\def\balpha{\mbox{\boldmath $\alpha$}}

\def\btau{\mbox{\boldmath $\tau$}}
\def\bnu{\mbox{\boldmath $\nu$}}
\def\bpsi{\mbox{\boldmath $\psi$}}

\def\bv{\mbox{\bf v}}

\input{amssymb.sty}

\def\frac#1#2{{{#1}\over{#2}}}
\def\tfrac#1#2{{\textstyle{{#1}\over{#2}}}}

\def\half{\tfrac{1}{2}}

\def\inner{\!\cdot\!}

\begin{document}

\begin{titlepage}

\baselineskip 24pt

\begin{center}

\parbox{13.82cm}{{\bf\Large Mass Hierarchy, Mixing, CP-Violation and Higgs Decay---or 
Why Rotation is Good for Us}}

\vspace{.5cm}

\baselineskip 14pt

{\large Michael J BAKER}\footnote{bakerm\,@\,maths.ox.ac.uk; {\it Mathematical Institute, University of Oxford,
  24-29 St. Giles', Oxford, OX1 3LB, United Kingdom}; work supported by an EPSRC Research Studentship.\vspace{0.2cm}}, 
{\large Jos\'e BORDES}\footnote{jose.m.bordes\,@\,uv.es; {\it Departament Fisica Teorica and IFIC, Centro Mixto CSIC, Universidad de Valencia,  calle Dr. Moliner 50, E-46100 Burjassot (Valencia), Spain}; work supported by Spanish MICINN under contract FPA2008-02878.\vspace{0.2cm}},\\
\vspace{0.4cm}
{\large CHAN Hong-Mo}\footnote{hong-mo.chan\,@\,stfc.ac.uk; {\it Rutherford Appleton Laboratory, Chilton, Didcot, Oxon, OX11 0QX, United Kingdom.}\vspace{0.2cm}},
{\large TSOU Sheung Tsun}\footnote{tsou\,@\,maths.ox.ac.uk; {\it Mathematical Institute, University of Oxford,
  24-29 St. Giles', Oxford, OX1 3LB, United Kingdom.}\vspace{0.2cm}}

\end{center}

\vspace{.3cm}

\begin{abstract}

The idea of a rank-one rotating mass matrix (R2M2) is reviewed 
detailing how it leads to ready explanations both for the fermion 
mass hierarchy and for the distinctive mixing patterns between up 
and down fermion states, which can be and have been tested against 
experiment and shown to be fully consistent with existing data.  
Further, R2M2 is seen to offer, as by-products: (i) a new solution 
of the strong CP problem in QCD by linking the theta-angle there 
to the Kobayashi-Maskawa CP-violating phase in the CKM matrix, and
(ii) some novel predictions of possible anomalies in Higgs decay 
observable in principle at the LHC.  A special effort is made to 
answer some questions raised.

\end{abstract}

\end{titlepage}

\clearpage

\baselineskip 14pt

\section{Old Story}

Although almost all the material in this first half is published
sometime ago, it is widely scattered in both time and space and
in parts difficult of access.  It is thus thought that collecting
it now together and presenting it anew with the clarity acquired from 
experience and hindsight would help the reader judge better its
significance.  It would help also for evaluating better the more 
recent results to be reviewed in the second half, and offers at 
the same time an opportunity to address in detail some questions
which might arise.

\subsection{Why R2M2?}

The need to inject from experiment the masses and mixing angles as 
inputs into the standard model subtracts much from our confidence
in it as a fundamental theory in spite of all its phenomenological
successes.  These not only account for some two-thirds of its many 
empirical parameters, but are themselves a little mysterious, showing 
a hierarchical pattern begging explanation which is not supplied, 
even qualitatively, in the standard model framework.  Presumably, 
the explanation will have to come elsewhere, perhaps in a more 
fundamental theory from which the standard model itself can be 
deduced as a consequence, and there is no lack of effort from the 
physics community to construct such a theory.  However,  before 
we succeed in doing so, it would be interesting to ask what sort of 
mechanism could give rise to such a hierarchical pattern so as 
to serve as a guideline for constructing our future theory.  The 
mechanism will have to be somewhat unusual, given that the fermions 
of different generations are seen in experiment to have almost 
identical interactions, pointing to a hidden symmetry among them, 
which is only mildly broken, if at all.  And yet the masses and 
mixing angles vary wildly from generation to generation, often by 
orders of magnitude.  Any perturbative breaking of the generation 
symmetry is unlikely to give such a mass and mixing pattern.  The
rank-one rotating mass matrix (R2M2) is a suggestion for filling
the gap.

The mass matrix obtained from the Yukawa term by 
substituting for the Higgs field its vacuum expectation value 
appears usually as:
\begin{equation}
m \half (1 + \gamma_5) + m^\dagger \half(1 - \gamma_5),
\label{massmat}
\end{equation}
but by a suitable relabelling of the $su(2)$-singlet right-handed
fermion fields, which cannot change the physics, one can recast it
in a hermitian form in which no $\gamma_5$ appears \cite{Weinberg2}.
Explicitly, any $m$ can always be diagonalized by unitary matrices
$U$ and $V$, thus $U m V^{-1} =$ diagonal matrix.  Then relabelling the
right-handed fields $\psi_R$ as $U^{-1} V \psi_R$ will give a new
mass matrix with the desired property.  We shall always, in what 
follows, work with the mass matrix in this representation, in which 
case a rank-one mass matrix will appear simply as just a product of 
the eigenvector $\balpha$ with the single nonzero eigenvalue times
its hermitian conjugate, thus:
\begin{equation}
m = m_T {\balpha}{\balpha}^{\dagger}.
\label{mfact}
\end{equation}

That one should perhaps start with a rank-one fermion mass matrix,
i.e. one reducible to the form (\ref{mfact}), where ${\balpha}$, a 
unit vector in 3-D generation space, is the same for both the up 
and down states and only the coefficient $m_T$ depends on flavour 
(as in $m_U$ and $m_D$), has been suggested already some thirty years 
ago \cite{Fritsch,Harari}.  This has only one massive generation and 
has the unit matrix as the mixing matrix, which is already not a bad 
first approximation, indeed tantalisingly close, to the experimental 
situation, at least for quarks.  The idea then is that the small 
lower generation masses and deviations of the mixing matrix from 
the identity will arise from some perturbation to the above initial
situation.  

The difficult question, of course, is what perturbative mechanism 
will give rise to the hierarchical pattern of deviations from this 
first approximation.  For example, in order to obtain non-trivial 
mixing between the up and down states, it would appear necessary 
at first sight to break the degeneracy of $\balpha$ in up-down
flavour.  But the only forces we know in nature which depend on
flavour come from the electroweak sector, and these seem too weak
to give the desired effect; strong interactions, on the other hand,
where one perhaps expects effects of such magnitudes to originate,
are however flavour-blind.  

That the mass matrix, or in the case of (\ref{mfact}) the vector 
${\balpha}$, should rotate is a solution suggested to resolve this 
dilemma.  By a rotating mass matrix, whether of rank one or not, we 
mean one the orientation of which in generation space depends on the 
renormalization scale.  In other words, its eigenvectors change with
scale. We are familiar already with mass matrices having eigenvalues 
which depend on scale, 
and hence give scale-dependent masses to the 
individual fermions \cite{qcdrun}, e.g.\ to the $b$-quark, which has 
even been 
tested experimentally.  So a rotating mass matrix is just a further 
extension of the same idea of scale-dependence.  The beauty of a 
rotating mass matrix is that even without breaking the degeneracy of 
$\balpha$ in flavour, mixing and mass hierarchy will automatically
emerge solely by virtue of the dependence of $\balpha$ on scale.  This
means that mixing of sufficient magnitude can now in principle arise
via rotation from strong interaction, which does not distinguish between
up and down flavours.  Of 
course, actually how it can come about there is itself an interesting
question which will be outlined only briefly in a later subsection, 
since it is one of some intricacy and its resolution model-dependent, 
so that the reader will have to be referred elsewhere for details.
Here, in this paper, R2M2 will be treated just as a phenomenological
hypothesis to be tested against experiment and, if proved successful, 
to serve as hints for the construction of future theories to bolster,
or perhaps eventually to supplant, the present standard model.

Now, for a rotating mass matrix, not only the masses but also the 
state vectors will depend on scale, which is a situation that one 
is not so familiar with and needs therefore more careful handling, 
giving results which may appear at first sight surprising.  One 
will find indeed that there are certain conceptions one takes for 
granted because of familiarity which will have to be given up and 
replaced.  One immediate example is precisely that it will give 
rise to the hierarchical mass and mixing patterns one seeks.

\begin{figure}
\centering
\input{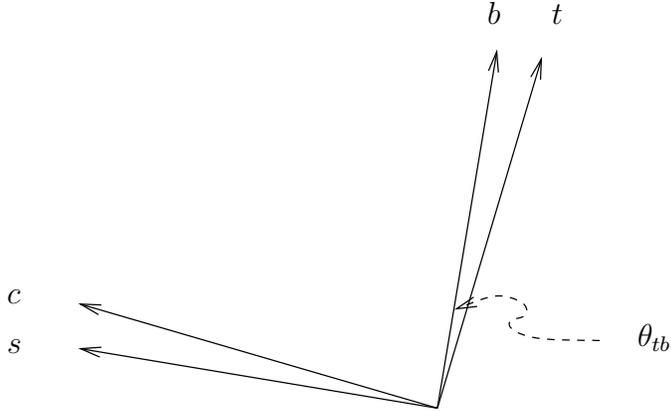}
\caption{Illustration of $UD$ mixing for two generations.}
\label{UDmix}
\end{figure}

That a rank-one rotating mass matrix can give rise to mixing and 
mass hierarchy is a very simple idea which is most easily seen in 
the simplified situation when there are only two generations 
instead of the three in reality, where one takes account only 
of the two heaviest states in each fermion type.  By (\ref{mfact}) 
then, taking for the moment ${\balpha}$ to be real and $m_T$ to 
be $\mu$-independent for simplicity, we would have $m_t = m_U$ as 
the mass of $t$ and the eigenvector ${\balpha}(\mu = m_t)$ as 
its state vector ${\bf v}_t$.  Similarly, we have $m_b = m_D$ as 
the mass and ${\balpha}(\mu = m_b)$ as the state vector 
${\bf v}_b$ of $b$.  Next, the state vector ${\bf v}_c$ of $c$ 
must be orthogonal to ${\bf v}_t$, $c$ being by definition an 
independent quantum state to $t$.  Similarly, the state vector 
${\bf v}_s$ of $s$ is orthogonal to ${\bf v}_b$.  So we have the 
situation as illustrated in Figure \ref{UDmix}, where the vectors 
${\bf v}_t$ and ${\bf v}_b$ are not aligned, being the vector 
${\balpha}(\mu)$ taken at two different values of its argument 
$\mu$, and ${\balpha}$ by assumption rotates. 
This gives then the following CKM mixing (sub)matrix in the situation 
considered with only the two heaviest states:
\begin{equation}
\left( \begin{array}{cc} V_{cs} & V_{cb} \\ V_{ts} & V_{tb} \end{array}
    \right) = \left( \begin{array}{cc} \langle {\bf v}_c|{\bf v}_s \rangle
                              & \langle {\bf v}_c|{\bf v}_b \rangle \\
                                \langle {\bf v}_t|{\bf v}_s \rangle
                              & \langle {\bf v}_t|{\bf v}_b \rangle
              \end{array} \right )
            = \left( \begin{array}{cc} \cos \theta_{tb} & -\sin \theta_{tb} \\
                 \sin \theta_{tb} & \cos \theta_{tb} \end{array} \right), 
\label{UDmix2}
\end{equation}
which is no longer the identity: hence mixing.

\begin{figure}
\centering
\input{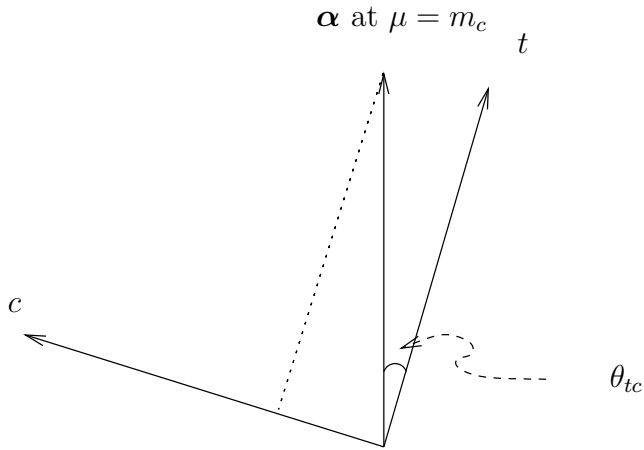}
\caption{Illustration of mass leakage for two generations.}
\label{alphaU}
\end{figure}

Next, what about hierarchical masses?  From (\ref{mfact}), it follows 
that ${\bf v}_c$ must have zero eigenvalue at $\mu = m_t$.  But this 
value is not to be taken as the mass of $c$ which has to be 
measured at $\mu = m_c$.  In other words, $m_c$ is instead to be taken 
as the solution to the equation:
\begin{equation}
\mu = \langle {\bf v}_c|m(\mu)|{\bf v}_c \rangle
     = m_U |\langle {\bf v}_c|{\balpha}(\mu) \rangle|^2.
\label{solvmc}
\end{equation}
A nonzero solution exists since the scale on the LHS decreases from 
$\mu=m_t$ while the RHS increases from zero at that scale.  Another
way to see this is that since ${\balpha}$ by assumption rotates
so that at $\mu < m_t$, it would have rotated to some direction
different from ${\bf v}_t$, as illustrated in Figure \ref{alphaU}, 
and acquired a component, say $\sin \theta _{tc}$, in the direction 
of ${\bf v}_c$ giving thus:
\begin{equation}
m_c = m_t \sin^2 \theta_{tc}.
\label{mc2}
\end{equation}
This is nonzero but will be small if the rotation is not too fast:
hence mass hierarchy. 

We see then that with 2 generations, rotation will give both mixing 
and mass hierarchy.  Basically the same argument can be applied also
to the realistic 3-generation case, but the situation becomes more
intricate and therefore more interesting, deserving a separate and
more detailed treatment later.  Before we do so, however, let us
first ask a few urgent questions triggered by the assertions above.

\subsection{Does rotation work?}

\begin{figure}
\centering
\hspace*{-3.5cm}
\input{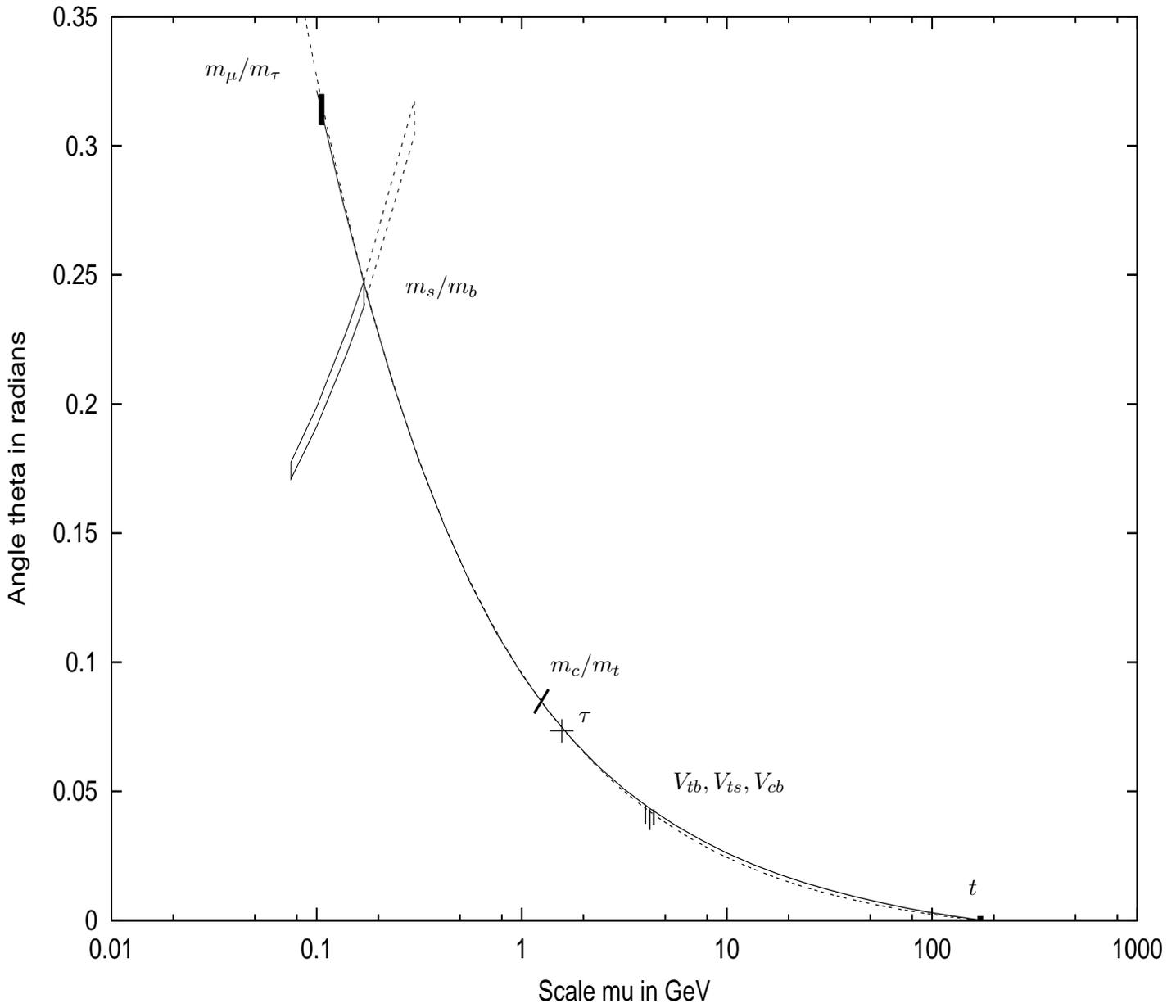}
\caption{The rotation angle changing with scale as extracted from the
  data on mass ratios and mixing angles, in planar approximation,
  taken from \cite{cevidsm}. }
\label{planarplot}
\end{figure}

The first is obviously the following.  Granted that R2M2 does give
rise to mass hierarchy and mixing, are the mass spectrum and 
mixing pattern anywhere near what is experimentally observed?  In
the planar approximation discussed in the preceding subsection, 
this question can easily be answered. As seen in
(\ref{UDmix2}) and (\ref{mc2}) the mixing angles and mass ratios 
are given explicitly in terms of the rotation angle.  Hence, given 
the experimentally measured values of these quantities, one can 
easily invert the formulae to evaluate the values of the rotation
angle to which they correspond.  Then, if R2M2 works, the rotation 
angles so obtained when plotted against the scales at which the
various mixing angles and masses are measured should all lie on a 
smooth curve representing the rotation trajectory of the vector
${\balpha}$ appearing in the mass matrix (\ref{mfact}).  This 
exercise was done in \cite{cevidsm} and is reproduced here in 
Figure \ref{planarplot}.  This clearly shows that the data 
then available were perfectly consistent with the R2M2 hypothesis 
in the ``planar approximation'' in which account is taken only of 
the 2 heaviest generations.

Whether the rotation mechanism will still work in the realistic
situation when all 3 fermion generations are taken into account
will be considered later when the 3-G analysis has been performed.
One can ask at this stage, however, why is it that
by missing out the lowest
generation as was done above, one still obtains a sensible picture.
For 3 generations (3-G), the vector ${\balpha}$ will trace out
for changing $\mu$ a curve on the unit sphere.  By restricting 
oneself to only the two heaviest generations, one is projecting
this space curve on to the plane containing the state vectors of
the two heaviest generations, and this ``planar approximation''
would be reasonable so long as the curve in the region studied is
nearly planar.  As will be seen later, the degree of nonplanarity
of the trajectory for ${\balpha}$ in the region considered is
measured by the Cabibbo angle $\sim 0.22$, and so long as one stays 
in that region, nonplanar effects are only of the order $0.22^2
\sim 0.04$ and are negligible to the accuracy needed \cite{cevidsm}.
Beyond the region shown in Figure \ref{planarplot}, i.e.\ for, say,
$\mu \leq m_\mu$, nonplanar effects will have to be taken into account,
as will be done when the time comes.

The exercise summarized in Figure \ref{planarplot}, apart from 
answering the question posed above, also gives some additional 
indications for rotation which were at first not bargained for.
\begin{description}
\item[(U)] In first proposing the formula (\ref{mfact}) for the
rotating mass matrix, one is guided by the experimental fact that
the CKM mixing matrix \cite{Cabibbo, KMphase}
is close to unity to insert the same vector
${\balpha}$ for the up and down quark states, as suggested a long 
time ago already \cite{Fritsch,Harari}.  But there is a priori no 
similar reason to suggest that the mass matrix for charged leptons 
should also have the same $\balpha$.  Nevertheless, one finds in 
Figure \ref{planarplot} that the masses for the charged leptons 
$\tau$ and $\mu$ actually fit very well on to the same trajectory, 
pointing to some sort of universality for ${\balpha}$.
\item[(F)] As seen in Figure \ref{planarplot}, the trajectory for
${\balpha}$ seems to approach an asymptote as $\mu \to
\infty$.  Indeed, the best empirical fit to the data points as
shown in the figure is an exponential, indicative of a rotational 
fixed point at $\mu = \infty$, meaning that the rotation will 
get progressively slower for larger $\mu$.  This can be seen also 
from the experimental values of the mass ratios: $m_\mu/m_\tau 
> m_s/m_b > m_c/m_t$ getting progressively smaller as the scale 
increases, since in the rotation scenario, it is the leakage by 
rotation which gives rise to the mass ratio of the second to the 
heaviest generations, and as the rotation rate gets smaller, so 
will the leakage.
\end{description}

These two observations will be designated in what follows as 
assumptions ({\bf U)} and ({\bf F)}.  Though initially quite 
independent of the R2M2 hypothesis, they will enhance by much 
its predictive power, and may be taken as additional hints for 
constructing the future theory behind the standard model.

\subsection{How does rotation arise?}

The R2M2 hypothesis has two components, namely that the fermion 
mass matrix is first, of rank one, and secondly, rotating with
changing scale. The two components are not contingent one on 
the other, for clearly the mass matrix can be of rank one without
rotating, or it can rotate without being of rank one.  Although 
the two components conspire in the application to give the 
explanation for mixing and mass hierarchy detailed above, in 
theoretical considerations, it is more convenient to separate 
the two so as to clarify the different issues involved.

That it seems expedient to start with a mass matrix of rank one
is an old idea \cite{Fritsch,Harari} with which one would be 
familiar with, but that it should rotate is a proposal not often
met with in the literature except in the present context, and it
would be appropriate, before we go any further, to ask first why, 
and under what conditions, would a mass matrix rotate.

The answer is easy to find, for even in the standard model as 
usually formulated, so long as there is nontrivial mixing between 
up and down fermion states, then rotation of the fermion mass 
matrices will occur.  This was pointed out already long ago in
e.g.\ \cite{Ramond}, although not often noted, and can easily be 
seen as follows.  The renormalization group equation for the mass 
matrices $U$ and $D$ of respectively the up and down quarks to 
leading order can be written as \cite{smrge}:
\begin{equation}
16 \pi^2 \frac{dU}{dt} = \frac{3}{2} (UU^\dag - DD^\dag)U
    +(\Sigma_U - A_U)U,
\label{rgeq1}
\end{equation}
\begin{equation}
16 \pi^2 \frac{dD}{dt} = \frac{3}{2} (DD^\dag - UU^\dag)D
    +(\Sigma_D - A_D)D,
\label{rgeq2}
\end{equation}
where in each equation the first term on the right represents the
contribution of the Higgs boson loop while the second term that 
of the gluon loop.  Of the two terms, what is of interest here is 
the former, where we recall that by nontrivial mixing between up 
and down states, we mean that the matrices $U$ and $D$ are related 
by a nondiagonal matrix $V$.  Thus $D = VUV^\dag$ or $U = V^\dag DV$, so 
that $U$ and $D$ cannot be simultaneously diagonalized.  Suppose now at 
some $t = \ln \mu$ we diagonalize $U$ in (\ref{rgeq1}) by a unitary 
transformation, which we can since $U$ is hermitian.  All terms in 
(\ref{rgeq1}) are then diagonal except $DD^\dag U$, which will thus
necessarily de-diagonalize the matrix $U$ on running by (\ref{rgeq1}) 
to a different $t$.  

In other words the matrix $U$ will rotate with changing $t$ or $\mu$, 
as asserted.  Similar arguments hold also for the rotation of the 
matrix $D$.  They will apply as well to the lepton mass matrices 
given that experiments on neutrino oscillation imply a nontrivial 
(PMNS) mixing \cite{pmns} also for leptons.  
The only reason why this fact is 
not more commonly noted is that the effect is small, given the 
smallness of the up-down mixing itself which is driving the rotation.

We conclude therefore that there is actually nothing very unusual
in the stipulation in R2M2 of a mass matrix which rotates with 
changing scale, and that rotation is connected to mixing,
since it occurs already in the standard model as
commonly formulated.  What is unusual is only that instead of the
up-down mixing giving rise to the rotation as demonstrated above 
for the standard model, one wants instead that it is the rotation 
of the fermion mass matrix which is giving rise to the up-down 
mixing.  This means two things: first, that the rotation has to be 
much faster than is generated by the above equations (\ref{rgeq1}) 
and (\ref{rgeq2}), and secondly, that it has to be driven by some 
other effects yet to be identified which lie outside the standard 
model framework.

Clearly, if that is the direction one wants to go, what is needed 
is some new renormalization effect which will give a term in the 
renormalization group equation that is not diagonal when the mass 
matrix is diagonal.  This looks like that one will have to look for 
some more fundamental theory which will give rise to such an effect
and is at the same time compatible with the standard model so as 
not to destroy its near-perfect agreement with experiment at present.  
And one will, presumably, want this theory also to have a reasonably 
appealing theoretical foundation.

An attempt along these lines \cite{prepsm,phenodsm,ckm},
suggested by ourselves and designated 
as the framed standard model (FSM), will be  reported elsewhere 
\cite{dfsm}.  This starts from the premises of what could be 
called a framed gauge theory (FGT) framework \cite{efgt} where in 
addition to 
the usual gauge boson and matter fermion fields of ordinary gauge 
theories, one introduces also the frame vectors (vielbeins) in 
internal symmetry space as dynamical variables (framons), as one 
does the vierbeins in gravity.  Just as the vierbeins in gravity,
the framons in FGT will carry two types of indices, referring 
one to the local, and the other to the global, reference frames,
and the action will have to be invariant under both the local and 
global symmetries.  The FSM appears in this FGT framework as the 
``minimally framed'' theory with the standard model gauge symmetry 
$su(3) \times su(2) \times u(1)$, by virtue of which fact, it has 
been shown to lead automatically to the following results:
\begin{itemize}
\item  It contains within its structure a scalar boson to play
the role of the usual Higgs field, as well as a global $su(3)$ 
symmetry to play the role of fermion generations.  
\item  It gives a fermion mass matrix of the factorizable form 
(\ref{mfact}) with ${\balpha}$ independent of the fermion type, 
i.e.\ satisfying assumption ({\bf U)} of \S1.2 above. 
\item  By virtue of the required invariance under both the local
and global symmetries, the vacuum is degenerate and is coupled to 
the vector $\balpha$ appearing in fermion mass matrix (\ref{mfact}).
\item  Under renormalization in the strong sector, the vacuum is
found to change with scale, and drags $\balpha$ along with it; 
hence rotation.
\item  This rotating $\balpha$ has a fixed point at infinite scale,
i.e.\ satisfying assumption ({\bf F)} of \S1.2 above.
\end{itemize} 
In other words, the FSM satisfies R2M2 and possesses the properties 
identified above as essential for R2M2 to reproduce fermion mixing 
and mass hierarchy, while keeping still the main features of the 
standard model.  At the same time, it is based on premises with, 
we think, a fair amount of theoretical appeal.

However, for the purpose of the present paper, the FSM serves only 
as an example that such rotational models are possible, while R2M2 
itself will continue to be treated as a phenomenological hypothesis.  
In other words considerations are given only to those consequences 
of R2M2 which are independent of what physical mechanism is driving 
the rotation, so as the better to serve as a guide for model 
builders who may be starting with very different ideas.

\subsection{Stumbling block?}

At this juncture two questions, basically the same, may be asked, 
and indeed are often asked, which need to be addressed before one 
proceeds any further.  Since they can be both posed and clarified 
already in the planar approximation of \S1.1 with only 2 generations, 
it will now be done below.

\begin{description}
\item[Q1.]  The mass matrix of (\ref{mfact}) at any $\mu$ is 
invariant under a chiral transformation on the state orthogonal to 
${\balpha}(\mu)$, in particular at $\mu = m_t$, under a chiral 
transformation in the state ${\bf v}_c$.  {\it Chiral invariance is 
associated with zero mass particles.}  Yet, it was claimed in \S1.1 
that $m_c$ is nonzero.  Why?
\item[Q2.]  The mass matrix $m$ in (\ref{mfact}) has one zero
eigenvalue at every $\mu$, in particular at $\mu = m_t$, this zero
eigenvalue has as eigenvector ${\bf v}_c$.  {\it The eigenvalues of 
a mass matrix are by definition the masses of particles.}  Yet it 
was claimed that $m_c$ is nonzero.  Why? 
\end{description}

The point raised in these questions, though accentuated by the fact 
that the mass matrix is of rank one, concerns mainly only the other 
component of the R2M2 hypothesis, namely that the mass matrix rotates 
with changing scale.  Let us therefore start here  with a rotating 
mass matrix, not necessarily of rank one so as not to confuse the 
issue, only returning at the end to the rank-one case to answer 
the questions posed.

Now, the common concepts concerning the mass matrix, such as those 
set out in italics in {\bf Q1} and {\bf Q2} above, were obtained 
at first on the premises where $m$ has no scale-dependence.  On 
extending the consideration to scale-dependent and then to rotating 
mass matrices, one ought therefore logically to re-examine whether 
the concepts we have gathered before still remain valid.  This is 
particularly true for the rotating mass matrix, which is, in a sense, 
uncharted waters needing some caution to negotiate, for although it 
occurs already in the standard model as noted in the preceding 
subsection, the effect there was small and has not engaged much 
attention.  It is thus proposed that one goes back to the beginning 
where the idea of a mass matrix was first conceived, then work back 
step by step to the presently desired rotating, rank-one case.  
Notice that such an analysis would in any case be needed theoretically, 
even if one is satisfied with the standard model as it stands, and 
eventually even needed
practically when experimental accuracy reaches a point 
where the previously ignored effects of rotation become measurable.  
Only the search now for an explanation of mixing and mass hierarchy 
via rotation has brought the necessity forward.

As already noted, the mass matrix when originally  conceived was a 
scale-independent quantity, being the value of the Yukawa term when 
the Higgs field appearing in it is given its vacuum expectation value. 
Then indeed, its eigenvalues are to be interpreted as the masses of 
the physical particles.  After that one learns that the mass matrix 
so defined actually depends on the renormalization scale.  One meets 
first with the situation when only the eigenvalues 
$m_1(\mu)$ and $m_2(\mu)$ of the mass matrix 
are scale-dependent, not its respective eigenvectors $\bv_1$ and $\bv_2$,
i.e.\ the non-rotating case, 
which is a fair approximation to the standard model as aleady noted.
Once the quantities involved are dependent on the scale, one would 
have to specify at what scales physical parameters are to be measured. 
The standard convention is that the mass of each individual particle 
is to be measured at the scale equal to its mass.  Thus, specializing 
for the moment to the $U$-type quarks for the case under consideration, 
when only the eigenvalues depend on scale, the mass of the $t$-quark 
is the scale at which the larger eigenvalue, say $m_1(\mu)$, is equal 
to the scale itself, i.e.\ the solution to the equation:
\begin{equation}
m_1({\mu}) = \mu;
\label{m1mu}
\end{equation}
in other words, we find both the $t$-mass and its state vector:
\begin{equation}
m_1(m_t)=m_t \quad {\rm and} \quad \bv_1=\bv_t.
\end{equation}
By orthogonality,
\begin{equation}
\bv_2=\bv_c.
\end{equation}

Similarly, the mass of the $c$-quark is defined as the scale at which 
the smaller of the two eigenvalue say $m_2(\mu)$ is equal in value to 
the scale itself, i.e.\ solution to the equation (\ref{m1mu}), only with 
2 instead of 1, so that we have 
\begin{equation}
m_2 (m_c) =m_c,
\end{equation}
which is not equal to $m_2(m_t)$.
This is by now familiar.  Notice, however, that at 
this stage we have already departed from the original assertion set 
out in italics in {\bf Q2} above that the physical masses of particles 
are the eigenvalues of ``the'' mass matrix.  The two physical masses 
of $t$ and $c$ are both indeed eigenvalues, but they are evaluated at 
different scales, meaning that they are two eigenvalues of two distinct 
matrices $m(\mu = m_t)$ and $m(\mu = m_c)$.  There is in fact no single 
matrix among the set $m(\mu)$ which has $m_t$ and $m_c$ as its two 
eigenvalues.  In other words, the designation of $m(\mu)$ as the mass 
matrix has become only nominal, for in no case are the eigenvalues of 
a single $m(\mu)$ at any given 
$\mu$ the physical masses of the particles $t$ and $c$.

Indeed, there is yet another sense in which the matrix $m(\mu)$ has 
departed from the idea of a mass matrix as originally conceived.  At 
any energy scale $m_c < \mu < m_t$, the state $t$ cannot appear as a 
physical state, and only the state $c$ remains physical.  So there
should not be a mass matrix at all but just a number representing the 
scale-dependent ``mass'' of the $c$-quark, with no matrix elements 
referring to the $t$ state at all.  Recall the analogous case of the  
analytic two-channel $S$-matrix \cite{smatrix}, say,
\begin{equation}
S = \left( \begin{array}{cc} S_{aa} & S_{ab} \\
                              S_{ba} & S_{bb} \end{array} \right).
\label{analyticS}
\end{equation}
It is a $2 \times 2$ matrix and exists for all energies $E$, but for 
$E_b < E < E_a$, where $E_a$ and $E_b$ denote respectively the energy 
thresholds for the channels $a$ and $b$, it is only nominally the 
$S$-matrix, because in that energy range, the channel $a$ becomes 
unphysical.  The physical $S$-matrix, which by unitarity should refer 
only to physical channels, is there given just by the element $S_{bb}$, 
obtained via a truncation of the matrix (\ref{analyticS}) by removing 
the rows and columns referring to the unphysical channel $a$.  It 
would thus seem that the same should hold for the scale-dependent 
mass matrix $m(\mu)$ under consideration.  For $m_c < \mu < m_t$, the 
matrix $m(\mu)$ is only nominally the mass matrix.  The physical mass 
matrix is obtained by truncating $m(\mu)$ by removing the rows and 
columns referring to the state $t$ leaving only a number referring 
to the space orthogonal to ${\bf v}_t$, i.e.\ in the direction of 
${\bf v}_c$.  That is, it ought to be the case in principle, but in 
practice, one sees that in the case being considered it does not 
really matter.  When the matrix $m(\mu)$ is diagonalized at any 
$\mu$ in the non-rotating case we are at present considering, it 
will remain diagonalized for all $\mu$.  Then truncating $m(\mu)$ 
as prescribed above will leave us just the number $m_2(\mu)$.  So 
the mass of $c$ is still to be obtained by solving the equation 
$m_2(\mu) = \mu$ as was suggested, a result we have now confirmed 
by a more careful analysis.

The interesting question now is what happens when this ``nominal'' mass 
matrix $m(\mu)$ changes not only its eigenvalues  $m_1(\mu)$ and $m_2(\mu)$,
but its eigenvectors $\bv_1(\mu)$ and $\bv_2(\mu)$ 
as well, as the scale changes.  In other words, 
what happens if $m(\mu)$ rotates?  As before, the $t$ mass is the 
solution to equation (\ref{m1mu}), giving both the mass and state
vector of $t$:
\begin{equation}
m_1(m_t)=m_t \quad {\rm and} \quad \bv_1 (m_t)=\bv_t.
\end{equation}
By orthogonality,
\begin{equation}
\bv_2(m_t)=\bv_c.
\end{equation}
In contrast to the nonrotating case, however, the truncation 
required by unitarity for scales $\mu < m_t$ is no longer quite so 
uneventful as before.  For $\mu < m_t$, 
what has to be projected out is the state $t$ which becomes unphysical. 
What remains as the physical mass matrix for $m_c < \mu < m_t$ is the 
submatrix in the subspace orthogonal to ${\bf v}_t$, 
\begin{equation}
\hat{m} (\mu) = \langle \bv_c |m(\mu)| \bv_c \rangle,
\end{equation}
which is just a number in this 
simple 2-generation model,   The physical mass of $c$ is the value of the 
scale $\mu$ when this number is equal to the scale itself, namely the 
solution to the first equation in (\ref{solvmc}) of \S1.1 above.  
Explicitly,
\begin{equation}
\hat{m} (m_c)=\langle \bv_c | m(m_c) | \bv_c \rangle = m_c,
\end{equation}
which is not equal to 
\begin{equation}
m_2 (m_c)= \langle \bv_2 (m_c) | m(m_c) | \bv_2 (m_c) \rangle,
\end{equation}
since
\begin{equation}
\bv_2(m_c) \not= \bv_2(m_t) \quad {\rm and} \quad \bv_c = \bv_2 (m_t),
\end{equation}
because
${\bf v}_2(\mu = m_c)$ is necessarily orthogonal 
to ${\bf v}_1(\mu = m_c)$, but not, as $m(\mu)$ rotates, orthogonal 
to ${\bf v}_t = {\bf v}_1(\mu = m_t)$ as ${\bf v}_c$ is.  We conclude 
therefore that the physical mass of $m_c$ of $c$ is not an eigenvalue 
of the ``nominal'' mass matrix $m(\mu)$ at $\mu = m_c$, nor indeed 
necessarily at any other value of $\mu$.

One may ask what if one ignores truncation and insists on taking the 
eigenvalue $m_2(\mu)$ at $\mu = m_c$ as the mass of the $c$ quark just
as one did above for the case when $m(\mu)$ was not rotating.  Then one
would have the eigenvector ${\bf v}_2(\mu = m_c)$ as the state vector
of $c$, which is orthogonal to the other eigenvector of $m(\mu)$ at 
$\mu = m_c$, namely ${\bf v}_1(\mu = m_c)$.  But this means that it
cannot be orthogonal to ${\bf v}_t$ which is ${\bf v}_1(\mu = m_t)$, 
and ${\bf v}_1(\mu = m_t) \neq{\bf v}_1(\mu = m_c)$ since the mass
matrix $m(\mu)$ by assumption rotates.  In other words, the state
vector ${\bf v}_c$ of $c$ would have a component in the direction of 
the state vector ${\bf v}_t$ of $t$, supposedly an independent quantum 
state to $c$, contradicting thus unitarity.

With this conclusion in hand, let us return to the questions {\bf Q1} and 
{\bf Q2} posed at the beginning of the subsection.  Specializing now to a 
rotating mass matrix of rank one, namely (\ref{mfact}), gives the 
value of $m_c$ via the second equation in (\ref{solvmc}) of \S1.1, 
which we have already remarked is nonzero.  But this is now seen to 
pose no problem since the conclusion is that physical masses of particles 
for a rotating scenario need not, and in general are not, eigenvalues 
of the rotating nominal mass matrix for non-leading generations. Hence 
$m_c$ need not be zero even when the nonleading eigenvalue is zero for 
$m(\mu)$ at all $\mu$.  This thus falsifies the assertion in italics 
in {\bf Q2} and answers at the same time that question.  It also answers 
the question {\bf Q1} since the chiral symmetry of the nominal mass matrix 
$m(\mu)$ appearing in the action will be maintained so long as this 
matrix has a zero eigenvalue, which it always has, but this is not 
contradicted by the physical $c$ mass $m_c$ being nonzero, since $m_c$ 
is not an eigenvalue of $m(\mu)$ at $\mu = m_c$ nor necessarily at 
any other value of $\mu$, falsifying thus the assertion in italics 
there.

\subsection{Three generations: the realistic case}

Having clarified in preceding subsections some immediate questions 
concerning the R2M2 hypothesis with the simplified 2-generation model, 
we are now ready to apply R2M2 to the realistic 3-generation case.

For three generations of fermions, the vector ${\balpha}$ in the 
rank-one mass matrix (\ref{mfact}) is a 3-vector, and as the scale 
$\mu$ varies, ${\balpha}$ traces out a trajectory on the unit 
sphere.  Now a space curve has of course more freedom than a plane 
curve, leading to more intricate features for both the mass spectrum 
and mixing pattern of fermions.  While a plane curve can bend only 
one way, a space curve on a surface can bend with the surface (normal 
curvature $\kappa_n$) or sideways (geodesic curvature $\kappa_g$), 
and it can also twist (torsion $\tau$) \cite{Docarmo}, but if the 
surface is the unit sphere, as in our case, then the relevant twist,
namely the geodesic
torsion $\tau_g$,
is zero.  This means that whereas the amount of bending achieved by a 
curve in either direction, i.e.\ whether with the surface or sideways, 
over a small displacement in arc-length $\delta s$ is proportional 
to $\delta s$, namely $\kappa_n \delta s$ or $\kappa_g \delta s$, any 
twist achieved over the same displacement will have to be of at least 
second order in $\delta s$.  As will be seen, this simple geometric 
fact will already go a long way towards explaining the hierarchical 
mass spectrum and mixing pattern of fermions observed in experiment.

Let us first recall here the experimental facts which we are meant 
to explain.  According to the PDG data \cite{databook}, 
the fermion mass 
spectrum is roughly as follows:
\begin{equation}
\begin{array}{llll}
    m_t \sim 171.3\ {\rm GeV} & m_b \sim 4.20\ {\rm GeV} & 
m_\tau \sim 1.777\ {\rm GeV} \\
    m_c \sim 1.27\ {\rm GeV} & m_s \sim 105\ {\rm MeV} & 
m_\mu \sim 106\ {\rm MeV} \\
    m_u \sim 2\ {\rm MeV} & m_d \sim 5\ {\rm MeV} 
& m_e \sim 0.5\ {\rm MeV}, \end{array}
\label{fmspectrum}
\end{equation}
and 
\begin{equation}
|\Delta m_{32}^2| \sim 2.43 \times 10^{-3} \ {\rm eV}^2; \ \ 
|\Delta m_{21}^2| \sim 7.59 \times 10^{-5} \ {\rm eV}^2
\label{numass}
\end{equation}
while the CKM mixing matrix elements (absolute values) for quarks 
appear as:
\begin{equation}
|V_{CKM}|  \sim \left( \begin{array}{lll}
0.97419 & 0.2257 & 0.00359 \\
0.2256 & 0.97334 & 0.0415 \\
0.00874 & 0.0407 & 0.999133 \end{array} \right), 
\label{CKM}
\end{equation}
and the PMNS mixing angles for leptons are given by:
\begin{equation}
\sin^2 \theta_{12} \sim 0.32, \ \sin^2 \theta_{13} < 0.05, \ 
\sin^2 \theta_{23} > 0.36.
\label{MNS}
\end{equation}

To explain the actual values of the mass and mixing parameters would 
be the aim of an eventual theory behind the standard model, but for 
the R2M2 hypothesis of our concern here the main interest is to first 
understand the qualitative patterns.

For these, in the mass spectrum, one can identify the following:

\begin{description}
\item[(A1)]  For each of the 3 fermion types $U$, $D$, and the
charged leptons $L$, the masses decrease by a large factor, each of 
an order of magnitude or more, from generation to generation.  For
neutrinos, although it is not yet clear whether their masses follow 
the same hierarchical pattern, existing data are fully consistent 
with it. 
\item[(A2)]  This factor, say between the second and the heaviest 
generations, changes from type to type, increasing with decreasing 
mass of the heaviest generation in the fermions type, thus: $m_c/m_t 
< m_s/m_b < m_\mu/m_\tau$, except again possibly for neutrinos.
\item[(A3)]  The masses in each generation also seem to change 
``hierarchically'' across the different types thus: $m_t \gg  m_b 
> m_\tau \gg m_{\nu_3}$, but here it is not clear what decides the 
ordering.  An exception is that $m_u < m_d$, bucking the trend, 
which is of course a crucial fact that guarantees the stability of 
the proton, and hence of us.
\end{description}

Whilst in the mixing matrices, one can identify the following:

\begin{description}
\item[(B1)]  The matrix elements for both quarks and leptons seem
to decrease hierarchically away from the diagonal, with the elements
at the corners (13 or 31) being at least about an order of magnitude
smaller than the others (12, 23, 21, 32). 
\item[(B2)]  The element (23) for quarks is much smaller than that
for leptons, while the element (12) remains about the same for both,
hence giving $V_{us}, V_{cd} \gg V_{cb}, V_{ts}$ for quarks.
\end{description}

We would like now to see whether the R2M2 hypothesis is 
able to explain these.

Let us examine first the mass hierarchy in a single fermion type,
say the $U$-type quarks $t, c, u$.  Extending the analysis for the 
2-G case above leading to the formula (\ref{solvmc}) for the $c$ mass 
to the present 3-G case, one obtains straightforwardly:
\begin{eqnarray}
m_t & = & m_U, \nonumber \\
m_c & = & m_U |{\balpha}_c\inner{\bf v}_c|^2, \nonumber \\
m_u & = & m_U |{\balpha}_u\inner{\bf v}_u|^2.
\label{hiermass}
\end{eqnarray}
where one has yet to specify what are the vectors ${\bf v}_c$ and 
${\bf v}_u$.  For the 2-G case, the assertion that the $c$ and $t$ 
are independent quantum states and hence should have orthogonal 
state vectors already specifies ${\bf v}_c$, given that ${\bf v}_t 
= {\balpha}(m_t)$.  Here in the 3-G case, however, the same 
assertion still leaves us the ambiguity of which vector on the plane 
orthogonal to ${\bf v}_t = {\balpha}(m_t)$ is to be taken as 
${\bf v}_c$, although, of course, once having decided on ${\bf v}_c$, 
then ${\bf v}_u$ is determined.

This ambiguity is resolved as follows.  For scales $\mu < m_t$, the 
state $t$ becomes unphysical, hence also the ``mass matrix'' $m$ in 
(\ref{mfact}).  According to the discussion in the preceding section, 
the  mass matrix for these scales 
is to be obtained from the nominal matrix $m$ by projecting out the
unphysical $t$-state into the subspace orthogonal to ${\bf v}_t$,
which gives for the present case:
\begin{equation}
\hat{m}_{ij} = m_U \langle {\bf v}_i|{\balpha}\rangle
    \langle {\balpha}|{\bf v}_j \rangle,
\label{mhat}
\end{equation}
where ${\bf v}_i,\ i = 1, 2$ are any two basis vectors chosen on the
said orthogonal subspace.  Notice that this truncation leaves the 
$2 \times 2$ mass matrix $\hat{m}$ for $\mu < m_t$ still 
of the factorized form as was the original $3 \times 3$
matrix $m$.  Explicitly:
\begin{equation}
\hat{m} = \hat{m}_U \hat{{\balpha}} \hat{{\balpha}}^\dagger,
\label{mhatfact}
\end{equation}
if we put:
\begin{equation}
\hat{m}_U = m_U  N
\label{mhatO}
\end{equation}
with
\begin{equation}
N = \sqrt{|\langle {\bf v}_1|{\balpha}\rangle|^2
           + |\langle {\bf v}_2|{\balpha}\rangle|^2},
\label{N}
\end{equation}
\begin{equation}
\hat{{\balpha}} = N^{-1} \left( \begin{array}{c}
    \langle {\bf v}_1|{\balpha}\rangle \\
    \langle {\bf v}_2|{\balpha}\rangle \end{array} \right),
\label{alphahat}
\end{equation} 
and $\hat{{\balpha}}$ as the 2-vector factorized from $\hat{m}$
in the subspace orthogonal to ${\bf v}_t$.  In other words, the 
situation is entirely analogous to the $3 \times 3$ mass matrix 
that one started with originally, only now in one lower dimension. 
One concludes therefore that, as for ${\bf v}_t$ in $m$ before, 
the state vector ${\bf v}_c$ of c should likewise be 
taken as the unique massive eigenstate of $\hat{m}$, namely 
$\hat{\balpha}(\mu)$, taken at scale $\mu = m_c$.  The state vector 
${\bf v}_u$ is then determined by orthogonality to both ${\bf v}_t$ 
and ${\bf v}_c$.  Explicitly, we have then the whole $U$-triad as:
\begin{eqnarray}
{\bf v}_t & = & {\balpha}(m_t); \nonumber \\
{\bf v}_c & = & {\bf v}_u \times {\bf v}_t; \nonumber \\
{\bf v}_u & = & \frac{{\balpha}(m_t) \times {\balpha}(m_c)}
    {|{\balpha}(m_t) \times {\balpha}(m_c)|}.
\label{Utriad}
\end{eqnarray}
Coupled with (\ref{hiermass}), these equations then give all the 
masses and state vectors of the $U$-type quarks.

That the mass spectrum of $U$-type quarks so obtained should be
hierarchical, i.e.\ $m_t \gg m_c \gg m_u$, can now be seen from the
earlier geometric observations as follows.  From the definition of
the $U$-triad in (\ref{Utriad}) above, one sees that the $c$ mass 
$m_c$ is associated with the bending (curvature) of the trajectory
while the $u$ mass is associated with the twist (geodesic torsion).  Then 
the fact cited before that the rotation trajectory
on a round sphere has no geodesic torsion 
implies that while  $\sqrt{m_c}$  is of first order in displacement 
along the trajectory, $\sqrt{m_u}$ is necessarily of second or 
higher order.

The same conclusion can be seen in perhaps more physical language.
The state vector ${\bf v}_c$ of $c$ is the (normalized) vector 
orthogonal to ${\bf v}_t$ lying on the plane containing both 
${\balpha}(\mu = m_t)$ and ${\balpha}(\mu = m_c)$.  Hence it 
soaks up all the leakage of mass acquired by rotation from 
$\mu = m_t$ to $\mu = m_c$, leaving ${\bf v}_u$, by definition 
orthogonal to ${\bf v}_c$, with a zero eigenvalue at $\mu = m_c$.
Any mass for the $u$ state will thus have to rely on leakage from 
$m_c$ by rotation from $\mu = m_c$ to $\mu = m_u$, just as the mass
of the $c$ state did on leakage from $m_t$.  For the same reason 
then that $m_c/m_t$ is small, so $m_u/m_c$ will be small also.

The same considerations will apply to all fermion types, namely, 
also to $D$-quarks, charged leptons and neutrinos giving each of
the types a hierarchical mass spectrum, except that for neutrinos
there may be complications due to an expected see-saw mechanism.
In other words, the mass hierarchy is now fully extended to all 
3 generations, explaining the observed pattern listed as {\bf (A1)} above. 
Further, as noted already before at the end of \S1.2, the 
R2M2 hypothesis when combined with the assumptions {\bf (U)} and 
{\bf (F)}, 
namely of respectively the universal trajectory and the existence of
a high scale fixed point, will explain also the pattern listed
above as {\bf (A2)}.  What is left unexplained are the points made in
{\bf (A3)}, where apart from the great mystery of $m_d>m_u$, no distinctive
pattern emerges among the flavour states.

Next, we turn to the mixing matrices and seek again a geometric
explanation for the patterns {\bf (B1)} and {\bf (B2)} observed in 
them above
in terms of the curvatures and torsion of the rotation trajectory.
Explicitly, the two curvatures are defined as follows.  Let us
first set up at each point on the trajectory a Darboux triad 
\cite{Docarmo} comprising in the case of a curve on the sphere 
(i) the radial vector ${\balpha}$ (ii) the tangent vector 
${\btau}$, (iii) the vector ${\bnu}$ orthogonal to both 
${\balpha}$ and ${\btau}$, all of unit length.  We take 
the tangent vector $\btau(\mu)$ and differentiate it with respect 
to the arc length $s$ along the trajectory.  This gives us the 
curvature vector, which must be orthogonal to $\balpha(\mu)$ given 
that $\balpha$ is of unit length.  The component of the curvature
vector in the direction of ${\balpha}$ is then the normal
curvature $\kappa_n$ while that in the direction ${\bnu}$ is
the geodesic curvature $\kappa_g$.  For a curve on the sphere,
the normal curvature is constant, being the curvature of the
sphere itself, but the geodesic curvature is a property of the 
curve and can depend also on the point on it.

To first order in $\delta s$, the Serret--Frenet--Darboux formulae 
\cite{Docarmo} relate two neighbouring points $\balpha$ and $\balpha'$ 
on the curve:
\begin{eqnarray}
\balpha'=\balpha(s+\delta s) & =& \balpha(s) - \kappa_n \btau (s)\,\delta s 
+ \tau_g \bnu(s)\,\delta s \nonumber\\
\btau'=\btau(s+\delta s) &=& \btau (s) + \kappa_n\balpha(s)\,\delta s
+ \kappa_g
\bnu (s)\,\delta s \nonumber \\
\bnu' = \bnu(s+\delta s) &=& \bnu(s) - \kappa_g \btau (s)\,\delta s -\tau_g
\balpha(s)\,\delta s
\label{sfd}
\end{eqnarray}
where we have kept the geodesic torsion term, which vanishes, to keep
track of which terms should be small.

From the formulae in (\ref{Utriad}), one readily sees that in the 
limit when the separation between $t$ and $c$ goes to zero, then 
${\bf v}_c$ becomes the tangent ${\btau}$ to the trajectory 
at $\mu = m_t$ while ${\bf v}_u$ becomes the normal ${\bnu}$, 
so that together with ${\bf v}_t = {\balpha}(\mu = m_t)$, they
constitute the Darboux triad at $\mu = m_t$.  Similarly, when 
the separation between $b$ and $s$ goes to zero, the 3 vectors
${\bf v}_b, {\bf v}_s, {\bf v}_d$ become the Darboux triad to the 
trajectory at $\mu = m_b$.  So if the separations between $t$
and $c$ and between $b$ and $s$ were both much smaller than the
separation between $t$ and $b$, then the CKM matrix would just be
the matrix relating the Darboux triads at the two neighbouring
points on the trajectory at $\mu = m_t$ and $\mu = m_b$.  To
first order in the separation $\delta s$ in arc length between
the two points, the Serret-Frenet-Darboux 
formulae (\ref{sfd}) give:
\begin{equation}
V_{CKM} \sim \left( \begin{array}{ccc}
\bnu \inner \bnu' & \bnu \inner \btau' & \bnu \inner \balpha'\\
\btau \inner \bnu' & \btau \inner \btau' & \btau \inner \balpha' \\
\balpha \inner \bnu' & \balpha \inner \btau' & \balpha \inner \balpha' 
\end{array} \right)
\sim \left( \begin{array}{ccc}
    1 &  \kappa_g \delta s &  \tau_g \delta s \\
    -\kappa_g  \delta s & 1 & -\kappa_n \delta s \\
    -\tau_g \delta s & \kappa_n \delta s & 1 \end{array} \right).
\label{ckmfromsfd}
\end{equation}
Since our curve is on the unit sphere, the geodesic 
torsion $\tau_g$ vanishes and the normal curvature $\kappa_n$ is 
unity.  It then immediately follows that the two corner elements 
$V_{ub}, V_{td}$, being of order at least $\delta s^2$, have to be 
much smaller than the other nondiagonal elements $V_{cb}, V_{ts}$ 
and $V_{us}, V_{cd}$ of order $\delta s$, giving thus the pattern 
{\bf (B1)} in the CKM matrix already noted.

In reality, of course, the separation between the two heaviest
generations is not small for either $U$ or $D$ compared with the
separation between $t$ and $b$, so that the approximations made 
in the paragraph above to derive (\ref{ckmfromsfd}) do not actually 
apply.  Nevertheless, it gives us a qualitatively correct picture 
for what happens to the CKM matrix elements in relation to the 
behaviour of the rotation trajectory.  In particular, it is now 
easy to see why the CKM elements in the last row and column are 
hierarchical in a manner so reminiscent of the mass hierarchy, a 
similarity often remarked on.  In the present language, say for $U$ 
quarks, if we interpret the separation $\delta s$ as the displacement 
from $t$ to $c$, and from $t$ to $u$ respectively, the relations 
(\ref{hiermass}) would give us:
\begin{eqnarray}
m_c &\sim& m_t\, |\balpha' \inner \btau|^2 \sim m_t (\delta s)^2
\nonumber\\
m_u &\sim& m_t\, |\balpha' \inner \bnu|^2 \sim 0
\label{mcufromsfd}
\end{eqnarray}
the last term being $0$ as the geodesic torsion vanishes.  In other
words, both the elements $V_{cb}, V_{ts}$ and the mass ratios 
$m_c/m_t,\ m_s/m_b$ come from the bending of the trajectory along 
the sphere (normal curvature) and hence are much larger than 
respectively the corner elements $V_{ub}, V_{td}$ and the mass 
ratios $m_u/m_t,\ m_d/m_b$ both of which can come only from a twist 
in the trajectory, necessarily of second or higher order in the 
separation since the trajectory has no geodesic torsion.

The same considerations applied to the leptons will show that in 
the PMNS matrix the corner element $U_{e3}$ governed by the twist 
of the trajectory will be much smaller than either of the elements 
$U_{e2}$ and $U_{\mu 3}$ governed by the bend of the trajectory, 
which is indeed as observed.  In other words, the pattern {\bf (B1)} is 
now fully explained.

Turning next to the 
pattern {\bf (B2)}, we observe that since the normal curvature here is 
constant as already noted, the corresponding elements $23, 32$ in 
the mixing matrices will just be proportional in magnitude to the
separation between the two states labelled by $2$ and $3$.  Hence,
given the much smaller mass ratio $m_{\nu_3}/m_\mu$ compared with 
$m_s/m_t,\ m_c/m_b$, it follows that $U_{\mu 3}$ in the PMNS 
matrix would be much larger than the corresponding elements $V_{ts}, 
V_{cb}$ in the CKM matrix.  The fact that the quarks, being located 
nearer the high scale fixed point postulated in the assumption {\bf (F)} 
above, will be moving more slowly along the trajectory than the 
leptons will only serve to enhance the above difference.  This then 
accounts for half of the pattern {\bf (B2)} listed.

As for the elements of the mixing matrices proportional to the
geodesic curvature $\kappa_g$, namely $V_{us}, V_{cd}$ in $V_{CKM}$
and $U_{e 2}$ in $U_{PMNS}$, they depend both on the trajectory 
and on the point on it, and cannot therefore be deduced simply from 
the rotation hypothesis.  One notes, however, that they can be very 
different both in magnitude and in scale-dependence from the mixing 
elements proportional to the normal curvature considered in the 
preceding paragraph, as they indeed seem to be, as observed in {\bf (B2)}.  
As to in what way they differ is model-dependent and will have to be 
left to model-builders.  An explicit example is given in \cite{efgt,
dfsm}.

One important property of the 3-G mixing matrix not yet mentioned
with no analogue in the 2-G case is of course the Kobayashi-Maskawa
CP-violating phase \cite{KMphase}.  In the treatment above of R2M2, 
the mass matrix
(\ref{mfact}) factorizes into a vector ${\balpha}$ which has been
taken tacitly as real, although this need not be the case.  If the 
elements of ${\balpha}$ are taken complex but their relative phases 
do not change with scale, then since only lengths and inner products 
of vectors are of physical interest, the results would be no different
from what were obtained above with real ${\balpha}$.  The mixing
matrices, in particular, will still have no KM CP-violating phase.
Only when the elements of ${\balpha}$ are allowed to have relative
phases changing with scale can one obtain a nonzero KM phase in the 
mixing matrix in the rotation scheme.  Such a scenario is not easy 
to implement, given no empirical indication of such phase changes 
in ${\balpha}$.  Besides, in all the models so far constructed, the 
vector ${\balpha}$ turns out, for reasons that are well understood,
always to have a constant phase, i.e.\ effectively real.  At this
stage of the presentation, therefore, the absence of a KM phase in 
the mixing matrix should be counted as a major failure of the R2M2 
hypothesis.  However, in the next section, it will be shown how the 
R2M2 hypothesis contains in itself a quite intriguing mechanism for 
removing this shortcoming while offering at the same time a solution 
to the long-standing strong CP problem in QCD.

\subsection{Does it work for 3-G too?}

It is seen in the preceding subsection that the R2M2 hypothesis when
supplemented by the assumptions {\bf (U)} and {\bf (F)} does seem to 
go quite a 
long way towards explaining the qualitative features of both the 
fermion mass spectrum and the quark and lepton mixing matrices.  One
needs however further to ask, as one did for the 2-G case, whether 
it really works in practice when applied to the physical system at 
hand.  One can address this question in two ways.  One can devise a
model trajectory with the appropriate features depending perhaps on 
some parameters, then try to fit experimental data with it and see 
whether one obtains a reasonable description.  Or else, one can
proceed in a more model-independent manner, as one did in the 2-G
case before in \S1.2, starting with the data and see whether they can 
be fitted with some sensible trajectory.

The first approach was first tried, indeed historically 
even before the
R2M2 hypothesis was properly formulated.  A phenomenological model
(DSM) \cite{ckm} was constructed based on some theoretical ideas which
led to a renormalizaton group equation for $\balpha$, with the
property that it possesses 
a rotational fixed point at
$\mu = \infty$, satisfying thus assumption {\bf (F)} as desired.  That
the trajectory is universal, assumption {\bf (U)}, was inserted as input
from experiment.  The resulting trajectory depends on 3 adjustable 
real parameter, explicitly a coupling parameter governing the speed 
of rotation with respect to scale, plus two integration constants.
These 3 parameters were varied to fit the then existing data 
\cite{databook98}. 
One obtained \cite{phenodsm} reasonable values for the mass
ratios, and
good agreement for all entries of
the CKM matrix, as well as those of the PMNS matrix, except for
the solar neutrino angle $U_{e2}$
depending on the geodesic curvature sensitive to the details of 
the rotation trajectory, and for the light quark masses the exact
interpretation of which is complicated by confinement. 
This was not bad for 3 parameters, and thought at first to
be a success of the DSM model.

It was soon realized, however, that the agreement with experiment 
obtained was not due so much to the validity of the model which, 
being intended to be merely phenomenological, contained some rather
arbitrary features, but rather to the general 
concept of rotation, namely what one now calls the R2M2 hypothesis. 
What is relevant for the present is thus not so much the premises 
on which this model (DSM) was built but the fact that with a model
trajectory possessing the envisaged features, one was able to obtain
already 
quite a sensible description of the then available data, which can 
thus be taken tentatively as evidence that R2M2 works even in the 
3-G case.

The more direct approach for testing the R2M2 hypothesis, starting
with the data to deduce the shape of the trajectory, as was done in
Figure \ref{planarplot} for the 2-G case, was also attempted previously
in \cite{cevidsm,pt}.  But, at that time, the data did not yield much 
more than what was contained already in Figure \ref{planarplot}, and 
one did not know how to take account of CP-violation.  With now
considerably more accurate data, and some improvements in technique
now available, the analysis will be repeated in \S2.2 below after 
the problem of the KM CP-violating phase is resolved so that the 
effects of this phase can be properly incorporated as they should be.
We anticipate the result \cite{btfit}
in stating that it is positive, meaning that
the R2M2 hypothesis is found to be still consistent with the existing 
data of much improved accuracy.

\section{New Shoots}

The situation as reviewed in the last section, though reasonably
satisfying in reproducing sensibly the effects one started out to 
explain, still needs to be viewed with caution in that the effects
``explained'' were known before one started, and so in constructing
the answer, namely the R2M2 hypothesis, as it is now called, one 
might have simply doctored the answer to fit the desired outcome.
To be more convinced of its validity, one can proceed in two ways.
One can go back to theory and seek a justification there for the 
R2M2 hypothesis, by devising, for example, a viable model behind 
the standard model as outlined above, which will produce the R2M2 
automatically as a consequence.  Or else, one can go forward and 
seek new premises on which to apply the R2M2 hypothesis so as to 
resolve problems or make predictions for which it was originally 
not devised.  An effort along the first direction is reported in 
\cite{prepsm,dfsm} as already mentioned.  Some efforts along the 
second direction are what will be reviewed below.

\subsection{A tale of two CP's}

For an example of a problem to attack using R2M2 outside its original
remit of mass hierarchy and mixing, we have long had our eyes on 
the strong CP problem in QCD.  To see why, let us first recall 
what is meant by the strong CP problem.

The ``strong CP problem'' is a long-standing problem in QCD, which 
has been with us for over forty years \cite{weinbergbook}.
There, gauge invariance in
colour allows in principle a term in the action of the form:
\begin{equation}
{\cal L}_{\theta} = - \frac{\theta}{64 \pi^2}
    \epsilon^{\mu \nu \rho \sigma} F_{\mu \nu} F_{\rho \sigma}
\label{Ltheta}
\end{equation}
for arbitrary $\theta$.  This term, which is of topological origin,
violates CP and can lead to a large electric dipole moment for the 
neutron.  Experimentally, however, the existing limit of the neutron 
dipole moment is now $2.7 \times 10^{-26}$ e cm \cite{edm} 
which means that the angle $\theta$ in (\ref{Ltheta}) has, for some 
unknown reason, to be of the order $3 \times 10^{-10}$ or less 
\cite{weinbergbook}.

The favourite theoretical suggestion for resolving this strong CP 
problem is the axion theory \cite{PecceiQuinn, Weinberg1}, 
but the axion has been looked for 
experimentally since soon after the problem was recognized but so
far has  not been found.  Hence, any new insight into the problem
should be more than welcome.

The theta-angle term (\ref{Ltheta}) above
is of the same form as the change in the Jacobian in the Feynman
integral when one performs a chiral transformation on a fermion
field appearing as an integration variable.  Hence, in a theory
with $N$ flavours of quarks, a chiral transformation, thus:
\begin{equation}
\psi \longrightarrow \exp(i \alpha \gamma_5) \psi
\label{chiraltrans}
\end{equation}
on each quark flavour will yield a theta-angle term modified to:
\begin{equation}
\theta \longrightarrow \theta + 2 \sum_{F} \alpha_F.
\label{modtheta}
\end{equation}
Clearly, this can be made to vanish by a judicious choice of 
$\alpha_F$.  Since a mere change of integration variables cannot
change the value of the integral which contains the physics, it
would seem that the effect of the theta-angle term can thereby be
eliminated.  However, the above chiral change of variables would 
also affect the integrand, in particular the mass term of the 
quarks if they are massive, and it will make their mass parameters
in general complex:
\begin{eqnarray}
  \!\!\! m \bar{\psi} \psi
    & = & m \bar{\psi} \half (1 + \gamma_5) \psi
    + m \bar{\psi} \half (1 - \gamma_5) \psi \nonumber \\
    & \rightarrow &
    m \exp(2i \alpha) \bar{\psi} \half (1 + \gamma_5) \psi
  + m \exp(- 2i \alpha) \bar{\psi} \half (1 - \gamma_5) \psi,
\label{modmass}
\end{eqnarray}
leading again to CP-violation.  Only when there is a quark with 
zero mass can the theta-angle term be eliminated without having to 
pay the penalty, a fact already recognized again for a long time 
as a possible solution to the strong CP problem.  Unfortunately, 
none of the quarks known can be assigned a zero mass in experiment, 
and so the problem remains.

The reason one thought right away \cite{Jakov,genmixdsm,strongcp} 
that R2M2 might be relevant for the strong CP problem is the fact
noted above in \S1.4 that R2M2 keeps chiral invariance
while allowing all quark masses to be nonzero.  This would seem to
mean that chiral transformations can be performed to eliminate 
any theta-angle term in the action and yet give non-zero masses
for all quarks as experiment seems to demand.  That observation is
however incomplete in that even in the usual scenario, the problem
is not with the elimination of the theta-angle term itself by a 
chiral transformation, which can always be performed, but with the 
effect of the transformation on the rest of the action, which can 
acquire thereby CP-violations elsewhere.  It would therefore be 
necessary in the R2M2 scenario also to examine how the rest of 
the action is affected by the chiral transformation, 
and whether or not the CP-violation is 
transmitted somewhere else.  To do so, one will need to perform 
the chiral transformation explicitly on the 3-component quark
field \bpsi\ (for 3 generations)
and then to follow through 
its effect down to the level of measured quantities in the R2M2 
scenario.

This has now been done as follows \cite{atof2cps}. One starts by 
considering first the R2M2 mass matrix (\ref{mfact}) at some chosen 
value of $\mu$.  It has two independent vectors with zero eigenvalues 
orthogonal to the rotating vector ${\balpha(\mu)}$, which may be 
taken as first, the normalized tangent vector ${\btau}(\mu)$ to the
trajectory which is orthogonal to ${\balpha(\mu)}$ since the
latter remains of unit length, and second, the normalized normal
vector ${\bnu}(\mu)$ orthogonal to both ${\balpha}(\mu)$ 
and ${\btau}(\mu)$.  These 3 vectors ${\balpha(\mu)}$,
${\btau}(\mu)$, and ${\bnu}(\mu)$, form 
the Darboux triad to the trajectory at $\mu$, 
as was noted already in \S1.5.  Now, both the vectors
${\btau}(\mu)$ and ${\bnu}(\mu)$ being eigenvectors of 
$m(\mu)$ with zero eigenvalues, a chiral transformation on 
the quark field corresponding to either component
can be performed to eliminate the theta-angle term at $\mu$, yet 
leaving $m(\mu)$ invariant and hence hermitian.  Consider however 
the mass matrix $m(\mu + d\mu)$ at a neighbouring point $\mu + d\mu$
on the trajectory, which is given explicitly in terms of 
$\dot{{\balpha}}(\mu)$ as:
\begin{equation}
m(\mu + d \mu) = m_T \left({\balpha}(\mu) + \dot{\balpha}(\mu)
    d\mu\right) \left({\balpha}(\mu) + \dot{\balpha}(\mu) 
    d\mu\right)^{\dagger}.
\label{mmu+dmu}
\end{equation}
One sees then that ${\btau}(\mu)$, which is proportional to
$\dot{{\balpha}}(\mu)$, is not an eigenvector of $m(\mu + d\mu)$ 
with zero eigenvalue, so that $m(\mu + d\mu)$ will not be left 
hermitian by a chiral transformation performed on \bpsi\ along the
component corresponding to the tangent ${\btau}(\mu)$.  Only a 
chiral transformation along the normal component ${\bnu}(\mu)$ at 
$\mu$ to eliminate the theta-angle term will leave both $m(\mu)$ 
and $m(\mu + d\mu)$ hermitian.

Iterating then the procedure, eliminating the theta-angle term at 
every $\mu$ by a chiral transformation on $\bnu \cdot \bpsi$, the 
component along the normal $\bnu(\mu)$ of the quark field, will 
leave the mass matrix $m(\mu)$ hermitian all along the trajectory.

To be explicit, let us choose a reference frame in 3-D generation 
space such that at $\mu = \infty$, we have:
\begin{equation}
{\balpha}(\infty) = {\balpha}_0 
   = \left( \begin{array}{c} 1 \\ 0 \\ 0 \end{array} \right); \ \
{\btau}(\infty) = {\btau}_0 
   = \left( \begin{array}{c} 0 \\ 1 \\ 0 \end{array} \right); \ \
{\bnu}(\infty) = {\bnu}_0 
   = \left( \begin{array}{c} 0 \\ 0 \\ 1 \end{array} \right),
\label{Darbouxinf}
\end{equation}
and define a rotation $A(\mu)$ such that:
\begin{equation}
{\balpha}(\mu) = A(\mu) {\balpha}_0; \ \
    {\btau}(\mu) = A(\mu) {\btau}_0; \ \
    {\bnu}(\mu) = A(\mu) {\bnu}_0.
\label{Amu}
\end{equation}
The chiral transformation needed to eliminate the theta-angle term
from the action at scale $\mu$ can then be represented as:
\begin{equation}
P(\mu) = A(\mu) P_0 A^{-1}(\mu), \ \ P_0 = \left( \begin{array}{ccc}
  1 & 0 & 0 \\ 0 & 1 & 0 \\ 0 & 0 & e^{- i \theta \gamma_5/2}
  \end{array} \right).
\label{Pmu}
\end{equation}
Effecting such a chiral transformation on the quark field at every
$\mu$ would then remove the theta-angle term entirely and leave the
QCD action CP-invariant and the mass matrix hermitian at every 
$\mu$.

Notice, however, that since ${\balpha}$ rotates, the normal
${\bnu}(\mu)$ will also change its direction as a function of
$\mu$.  Hence, to iterate the procedure at $\mu + d \mu$ so as
to eliminate the theta-angle term there while ensuring 
the hermiticity of the mass matrix at the next neighbouring point 
too, one has first to undo the previous chiral 
transformation performed at $\mu$ before, and then effect again 
the chiral transformation at $\mu + d \mu$, namely to apply the 
operator:
\begin{equation}
P(\mu + d \mu) P^{-1}(\mu) 
\label{paratrans}
\end{equation}
to the quark fields, in order to obtain the desired result.  This 
operator (\ref{paratrans}) acts thus as a kind of parallel transport, 
detailing effectively what is meant by the same or parallel (chiral) 
phases at two neighbouring points along the trajectory, and hence, 
by iteration, at any two points a finite distance apart.

To see how these chiral transformations will affect the conclusions
above of R2M2 on quark masses and mixing, let us start first with 
just one type of quarks, say, the $U$-type quarks, i.e.\ $t, c, u$. 
We shall label the Darboux triad $(\balpha, \btau, \bnu)$ at $\mu=m_t$
as $(\balpha_U, \btau_U, \bnu_U)$.
The state vector of $t$, i.e.\ ${\bf v}_t$, or just ${\bf t}$ for 
short, is defined as $\balpha_U$ and the state vectors 
${\bf c}, {\bf u}$ are to be orthogonal to it and are themselves 
mutually orthogonal.  It follows therefore that the dyad ${\bf c},
{\bf u}$ is related to the dyad $\btau_U,\bnu_U$
by a rotation in the plane orthogonal to ${\bf t}$, thus:
\begin{equation}
{\bf c} = \Omega_U {\btau}_U; \ \ {\bf u} = \Omega_U {\bnu}_U,
\label{taunutu}
\end{equation}
with 
\begin{equation}
\Omega_U = \left( \begin{array}{ccc} 1 & 0 & 0 \\
                   0 & \cos \omega_U & - \sin \omega_U \\
                   0 & \sin \omega_U & \cos \omega_U 
                  \end{array} \right),
\label{Omega}
\end{equation} 
$\omega_U$ being the angle between ${\bf c}$ and ${\btau}_U$.
This angle is small but nonzero since, according to our prescription
above, ${\bf c} = {\bf v}_c$ is the vector which is orthogonal to
${\balpha}(\mu = m_t)$ and lies on the plane containing both the 
vectors ${\balpha}(\mu = m_t)$ and ${\balpha}(\mu = m_c)$, 
while ${\btau}_U$ is the tangent to the trajectory at $\mu = m_t$; 
it is thus a measure of how much ${\balpha}(\mu)$ has rotated from
$\mu = m_t$ to $\mu = m_c$.  Suppose we wish again to evaluate the
mass of the $c$ quark in the rotation scenario as we did before but 
incorporating now the above procedure for eliminating the theta-angle
term, we shall have to take 
the $c$ quark field  $\psi_c (\mu=m_t)=
{\bf c} \cdot P(m_t) \bpsi$, but it will now have to be parallelly
transported by (\ref{paratrans}) to $\mu=m_c$, giving $\psi_c
(\mu=m_c) = {\bf c} \cdot P(m_c) \bpsi$.  The mass term also,
according to (\ref{modmass}) above, will now appear at $\mu = m_c$ 
as:
\begin{equation}
m_T \bar{\bpsi} P(m_c) \balpha (m_c) \balpha^\dagger (m_c) P(m_c)
\bpsi,
\label{modMc}
\end{equation}
where the operators $P(m_c)$ can in fact be omitted since the vector
$\balpha (m_c)$ is invariant under $P(m_c)$.  What interests us here
as far as the $c$ mass is concerned, according to the analysis in, for
example, \cite{strongcp}, is the diagonal contribution from the $c$
quark, namely:
\begin{eqnarray}
\lefteqn{
m_T  \bar{\bpsi} P(m_c) {\bf c} {\bf c}^\dagger \balpha (m_c)
\balpha^\dagger (m_c) {\bf c} {\bf c}^\dagger P(m_c) \bpsi} \nonumber \\
  &&= m_T | {\bf c} \cdot \balpha (m_c)|^2 \bar{\psi}_c (\mu=m_c)
\psi_c (\mu=m_c),
\label{ccontrm}
\end{eqnarray}
giving then the $c$ mass as:
\begin{equation}
m_c=m_T | {\bf c} \cdot  \balpha (m_c)|^2,
\label{mcagain}
\end{equation}
i.e., exactly the same as before (\ref{hiermass})
when no consideration was given to
the elimination of the theta-angle term.
The same conclusions will apply 
also to $m_u$.  One sees therefore that so long as there is only one 
type of quarks, one can always manage, with a rotating factorizable 
mass matrix, to eliminate any theta-angle term so as to maintain 
CP-conservation, while keeping the mass matrix hermitian, and having 
at the same time hierarchical but nonzero masses for all the quarks.

What happens, however, when there are both up-type and down-type 
quarks?  In that case, the two types can be coupled by the weak 
current via the CKM mixing matrix, and one has again to follow through 
the preceding arguments and trace out the consequence of eliminating 
the theta-angle term.  To do so, let us denote the state vectors of 
the $U$-type quarks defined above at $\mu = m_t$ together as:
\begin{equation}
V_U = ({\bf t}, {\bf c}, {\bf u}) = \left( \begin{array}{ccc}
    t_1 & c_1 & u_1 \\ t_2 & c_2 & u_2 \\ t_3 & c_3 & u_3
    \end{array} \right),
\label{VU} 
\end{equation}
and similarly the state vectors of the $D$-type quarks defined at
$\mu = m_b$ as:
\begin{equation}
V_D = ({\bf b}, {\bf s}, {\bf d}) = \left( \begin{array}{ccc}
    b_1 & s_1 & d_1 \\ b_2 & s_2 & d_2 \\ b_3 & s_3 & d_3
    \end{array} \right),
\label{VD} 
\end{equation}
where in the notation introduced above, one has from rotation:
\begin{eqnarray}
V_U & = & A_U \Omega_U; \ \  A_U = A(\mu = m_t); \\ \nonumber
V_D & = & A_D \Omega_D; \ \  A_D = A(\mu = m_b).
\label{VUVD}
\end{eqnarray}
What is of interest is the relative orientation of $V_U$ and $V_D$,
the matrix of inner products between the state vectors of the
$U$-type and $D$-type quarks being the starting point for 
the CKM mixing matrix we seek.  In
order to compare the orientation of the state vectors of one type
to those of the other, the two types being defined as they are at
two different scales, one needs first to parallelly transport the 
(chiral) phase of each to the same scale, say $X$, 
before one can make due comparison, thus:
\begin{eqnarray}
P_XP_U^{-1} V_U & = & P_X(A_U P_0^{-1} A_U^{-1})(A_U \Omega_U)
    = \tilde{V}_U; \\ \nonumber
P_XP_D^{-1} V_D & = & P_X(A_D P_0^{-1} A_D^{-1})(A_D \Omega_D)
    = \tilde{V}_D.
\label{VIVDtilde}
\end{eqnarray}
Hence we obtain the CKM matrix in this scenario as:
\begin{equation}
V_{CKM} = \tilde{V}_U^{-1} \tilde{V}_D
         = (\Omega_U^{-1} P_0 \Omega_U) V_U^{-1} V_D
           (\Omega_D^{-1} P_0^{-1} \Omega_D),
\label{VCKM}
\end{equation}
independent of the chosen scale $X$, as expected.
Since only the left-handed quark fields are involved in the weak
current, the chiral phase $\exp -i \theta \gamma_5/2$ in $P_0$ 
has been replaced by the phase factor $\exp -i \theta/2$.

Notice that  the factor $V_U^{-1} V_D$ would be the CKM matrix if
there were no theta-angle term to contend with, and it 
would be a real matrix with no CP-violating phase if we started with a 
real ${\balpha}$ as we have done.  It will be convenient to refer to
it as the $UD$ matrix.  By 
insisting on the chiral transformations to eliminate the theta-angle 
term throughout, one has injected some new phases into the 
mixing matrix elements, and hence the possibility of CP-violation,
which will indeed be the case if the phases introduced by the said 
chiral transformations cannot be removed by any changes in phase of 
the physical quark states.  Indeed, if we were to put both 
$\omega_U = 0$ and the corresponding $D$-type angle $\omega_D = 0$
in (\ref{VCKM}), one would have obtained vanishing values for the 
Jarlskog invariant \cite{Jarlskog}
and no CP-violation.
The reason is clear, since 
in that case the vector ${\bf u}$ would coincide with the normal 
vector ${\bnu}$ at $\mu = m_t$, the component of the quark field 
on which the chiral transformation 
is performed, and the effect on the CKM matrix would be the same as 
that of changing the phase of the physical $u$ field, which is 
arbitrary; and similarly for the phase of the physical $d$ field. 
If one were to calculate the Jarlskog invariant 
from (\ref{VCKM}) in this case, the phases would cancel and one 
obtains a zero value.  Since, however, $\omega_U$ and 
$\omega_D$ are nonzero by virtue 
of the rotation as explained above, this cancellation has now no 
reason to occur and one has in general a nonvanishing Jarlskog 
invariant and CP-violations as the result.

One concludes therefore that even if starting with a real vector
${\balpha}$ in the factorized mass matrix (\ref{mfact}) in R2M2,
the presence of a theta-angle term in the action and the process of
eliminating it through a chiral transformation as detailed above
will automatically give rise to CP-violations in the CKM matrix. 
But will this yield Jarlskog invariants and CP-violating effects of
the order observed in experiment?  We recall that the strong CP
angle $\theta$ from which this effect is supposed to originate can
take in principle any arbitrary value and so should be taken 
without prejudice as of
order unity, whereas the measured value of the Jarlskog invariant
is of order $3 \times 10^{-5}$, so that a suppression by about 4 
orders of magnitude is required in the process of transmitting the
CP-violating effects from the strong sector to the weak sector via 
rotation.  This is possible, so long as the rotation is relatively 
slow as is envisaged.  To see whether it is indeed the case, one 
can evaluate the Jarlskog invariant for (\ref{VCKM}) with, for 
example, its $2 \times 2$ submatrix labelled by the 2 heaviest 
states $t, c$ and $b, s$.  One obtains then an explicit expression 
for $J$ in terms of $\theta$, $\omega_U$, $\omega_D$ and elements of 
the $UD$ matrix $V_U^{-1} V_D$.  The angles $\omega_U$ and $\omega_D$, 
one has already noted to be of order $\epsilon$, the angle rotated 
by the vector ${\balpha}$ from the scale of the heaviest to that
of the second generation.  Further, from an earlier analysis of the
rotation picture \cite{features}, as reviewed in \S1.5
above, one has learned that the CKM matrix elements $V_{ts}, V_{cb}, 
V_{cd}, V_{us}$ proportional to the curvatures of the rotation 
trajectory are all of order $\epsilon$, while the corner elements 
$V_{td}, V_{ub}$ proportional to its geodesic torsion are of order at least 
$\epsilon^2$.  This is not to say, of course, that all four elements 
of order $\epsilon$ need be of the the same size for, as already 
remarked before, the two curvatures, normal and geodesic, can have 
quite different values.  But, for the order-of-magnitude estimate 
generic to the rotation scheme aimed for at the moment, one can 
ignore such details specific to a particular trajectory, and just
substitute the above estimates into the formula for $J$.  One finds 
then that $J$ is of order $\epsilon^4$ and proportional to 
$\sin (\theta/2)$.  An estimate for the value of $\epsilon$ can be 
obtained from the rotation formula (\ref{mcagain}) for the mass ratio 
of the second generation to the heaviest, leading to $\epsilon 
\sim 0.08$ for $m_c/m_t$, and $\epsilon \sim 0.15$ for $m_s/m_b$. 
This then gives an order-of-magnitude estimate for the Jarlskog 
invariant as:
\begin{equation}
J \sim \sin (\theta/2) \times 10^{-4},
\label{Jorder}
\end{equation}
which is quite consistent with the experimentally measured value 
\cite{databook} of $\sim 3 \times 10^{-5}$ for a strong CP angle 
$\theta$ of order unity.\footnote{In the above analysis, the vector 
${\balpha}$ is all along tacitly taken to be real, whereas it can, 
in principle, be complex and, as already noted in \S1.5, if the 
relative phases of its components vary with scale, then they can 
lead to CP-violating phases in the CKM matrix as well.  However, 
now that it is seen that even for a real ${\balpha}$ CP-violations 
can result from a theta-angle term, the other possibility has lost 
much of its appeal.}

One knows of course, whether in the rotation framework or otherwise, 
that once given the small values observed in experiment for mixing 
angles involving the two heaviest states $t$ and $b$, it will follow 
already that the CP-violating effects of the KM phase will be small, 
since it is known that for two generations there is no CP-violation,
which can thus arise only through mixing with $t$ and $b$.  A priori,
however, one can give no actual estimate for the size of the effects, 
not knowing how or where the KM phase originates.  The difference 
with the scheme here is that, first, having traced the origin of 
the KM phase via rotation back to the strong sector, one can give 
now an actual estimate for $J$, and second, since the rotation 
relates also the mixing angles of fermions to their hierarchical 
masses, as explained in \S1.3 above, the estimate can be derived 
with only mass ratios as inputs, and no knowledge of the mixing 
angles being required at all.  In other words, the rotation has 
indeed managed by itself, in transmitting the CP-violating effects 
from the strong to the weak sector, the suppression by some four 
orders of magnitude noted above as needed to bring them correctly
down to the order observed in experiment.

If one is willing to supply more empirical information such as the 
experimentally measured values of the mixing angles as inputs to 
the rotation scheme, then one would obtain more accurate estimates 
of the Jarlskog invariant from an assumed value of order unity for
the theta-angle, or conversely an estimate of the theta-angle from
the measured value of the Jarlskog invariant.  This will be done in 
the following subsection.  Before we do so, however, let us first 
pause for a moment to take stock of the new situation.

As far as the R2M2 hypothesis is concerned, the result represents,
first, the removal of a major shortcoming in the R2M2 framework as 
reviewed in the last section and noted at the end of \S1.5, which 
is significant for the further development of the framework.  And 
secondly, but equally significantly, it provides a first example of 
the R2M2 idea applied, successfully it seems, to problems outside 
the area for which it was originally devised.  We recall that the
R2M2 hypothesis was originally conceived as a means for explaining 
fermion mass hierarchy and mixing.  Here it has been applied to 
yield a solution for the long-standing strong-CP problem in QCD, 
which has at first sight little to do with the mass spectrum and 
mixing patterns of fermion states.  And, as far as one can see, no 
new assumption has been made in deriving the new 
result.  It is, therefore, in a sense a prediction, and adds much
to the credibility of the R2M2 hypothesis.  Thirdly, this result, 
relying as it does on the unusual property of R2M2 in giving nonzero 
quark masses despite maintaining strict chiral invariance in the 
action, opens the door to other problems of a similar sort.  In 
particular, one has in mind the problem of chiral symmetry breaking 
in QCD, for which R2M2 would now suggest a very novel picture, namely, 
that chiral symmetry is never really broken, the deviations from 
symmetry one sees being only due to the physical fermion states 
being wrongly identified as eigenstates of the nominal mass matrix 
derived from the chiral invariant action rather than correctly as 
eigenstates of the physical mass matrix obtained by a truncation 
thereof, as detailed in \S1.4.  The question of whether this picture 
for chiral symmetry breaking can hold or not is being worked on.  
If answered in the affirmative, it would 
obviously add further to the credibility of R2M2.

Furthermore, this new result on CP, in linking the theta-angle term
in the QCD action to the CP-violating phase in the CKM matrix, 
brings out a point of considerable theoretical interest way 
beyond the original phenomenological remits of the R2M2 hypothesis.  
Indeed, if the above result were to be adopted, it would change 
our conception of the CP problem altogether.  It used to be that 
one has, on the one hand, a strong CP problem with its origin in 
topology that one does not quite know how to solve and, on the 
other, what one may call a weak CP problem of the KM phase which 
one knows ought to be there but not whence its origin nor how to 
estimate its size.  The two sides were separate and apparently 
totally unrelated.  But now, what R2M2 seems to be telling us is 
that the two CP problems, strong and weak, are actually one and 
the same phenomenon.  The unwanted theta-angle term in the QCD
action, which invariance principles tell us ought to be there, can 
be eliminated by a chiral transformation of integration variables, 
or in other words by just a redefinition of what the CP operation 
means, at no other cost, it seems, than introducing a KM phase in 
the CKM matrix, which one would want in any case.  This means, of 
course, that the KM phase is now traced to a topological origin, 
i.e.\ same as the theta-angle term, and the relation between the 
two will yield an estimate for its size, as it seems correctly to 
have done.  Turning the argument around, this means also that the 
existence of the KM phase as observed in experiment is telling us, 
via rotation and the theta-angle term, about the topology of the 
QCD world, and can even give us an estimate of the value of the 
theta-angle, which, as seen, will be of order unity.  Even apart 
from the phenomenology, this seems altogether a quite enchanting 
theoretical picture to have emerged.

\subsection{Sequel to ``Does it work for 3-G too'?''}

Having now understood how the KM CP-violating phase is to appear
in the R2M2 scheme, which was the missing link in \S1.6, we are 
now in a position to re-examine the question whether the recent 
data of accuracy much superior to those used in earlier attempts
\cite{cevidsm} are still consistent with the R2M2 hypothesis.  In
other words, in the same spirit as Figure \ref{planarplot} in the 
planar approximation of \S1.2, we wish to know whether the new 
full 3-generation data can be fitted with a smooth trajectory for 
$\balpha(\mu)$.

The R2M2 scheme has $\boldsymbol{\alpha}$ as a 
fundamental object, and the state vectors are derived from this.
To test the R2M2 hypothesis, however, we shall start from real orthonormal state
vectors and use experimental data to find a consistent trajectory of 
$\balpha$.  We first choose the $U$-type quarks to have state vectors:

\begin{align}
\mathbf{v}_{u}& = (1,\ 0,\ 0)^\dagger,\\
\mathbf{v}_{c}& = (0,\ 1,\ 0)^\dagger,\\
\mathbf{v}_{t}& = (0,\ 0,\ 1)^\dagger,
\end{align}
which we are free to do, and which are real as explained above.  Once
we have chosen a $UD$ matrix we can use it to define the $D$-type quark
state vectors.  
There are no explicit physical constraints on the $UD$ matrix so we are free to choose any orthogonal matrix.

From the mass leakage mechanism, $\balpha(\mu=m_t) = \mathbf{v}_{t}$, and similarly $\balpha(\mu=m_b) = \mathbf{v}_{b}$.  The leakage mechanism then fixes $\balpha_{c}$ and $\balpha_{s}$ in terms of the quark masses:

\begin{align}
\balpha_{c}& = \sqrt{m_{c}/m_{t}}\  \mathbf{v}_{c} +  \sqrt{1 - m_{c}/m_{t}}\  \mathbf{v}_{t},\\
\balpha_{s} &= \sqrt{m_{s}/m_{b}}\  \mathbf{v}_{s} +  \sqrt{1 - m_{s}/m_{b}}\  \mathbf{v}_{b}.
\end{align}
The two vectors $\mathbf{v}_{t}$ and $\mathbf{v}_{c}$ define a plane.  All that the mass ratio $m_u/m_t$ tells us about $\balpha_{u}$ is the angle which it makes with this plane.  We can thus restrict $\balpha_{u}$, and similarly $\balpha_{d}$, to lie somewhere on a line, parametrized by $t \in [0,\ 2\pi)$:

\begin{align}
\balpha_{u}& = \sqrt{m_{u}/m_{t}}\ \mathbf{v}_u + \sqrt{1 - m_{u}/m_{t}}\  \mathbf{v}_c\cos (t) + \sqrt{1 - m_{u}/m_{t}}\ \mathbf{v}_t \sin (t),\\
\balpha_{d}& = \sqrt{m_{d}/m_{b}}\ \mathbf{v}_d + \sqrt{1 - m_{d}/m_{b}}\  \mathbf{v}_s\cos (t) + \sqrt{1 - m_{d}/m_{b}}\ \mathbf{v}_b \sin (t),
\end{align}
where we will choose the signs of the square roots so that $\balpha_{u,d}$ has a positive projection onto $\mathbf{v}_{u,d}$.  The restrictions on $\balpha_{i}$ for the charged leptons were found in an analogous way, replacing $\left( u, c, t\right)$ with $\left( e, \mu, \tau\right)$.  The neutrinos place only very weak restrictions on the trajectory of $\balpha$ since their masses, and the PMNS matrix elements, are not well measured.

\begin{figure}
\begin{center}
\begin{tabular}{cc}
\hspace{-0cm}
\includegraphics[height=6cm]{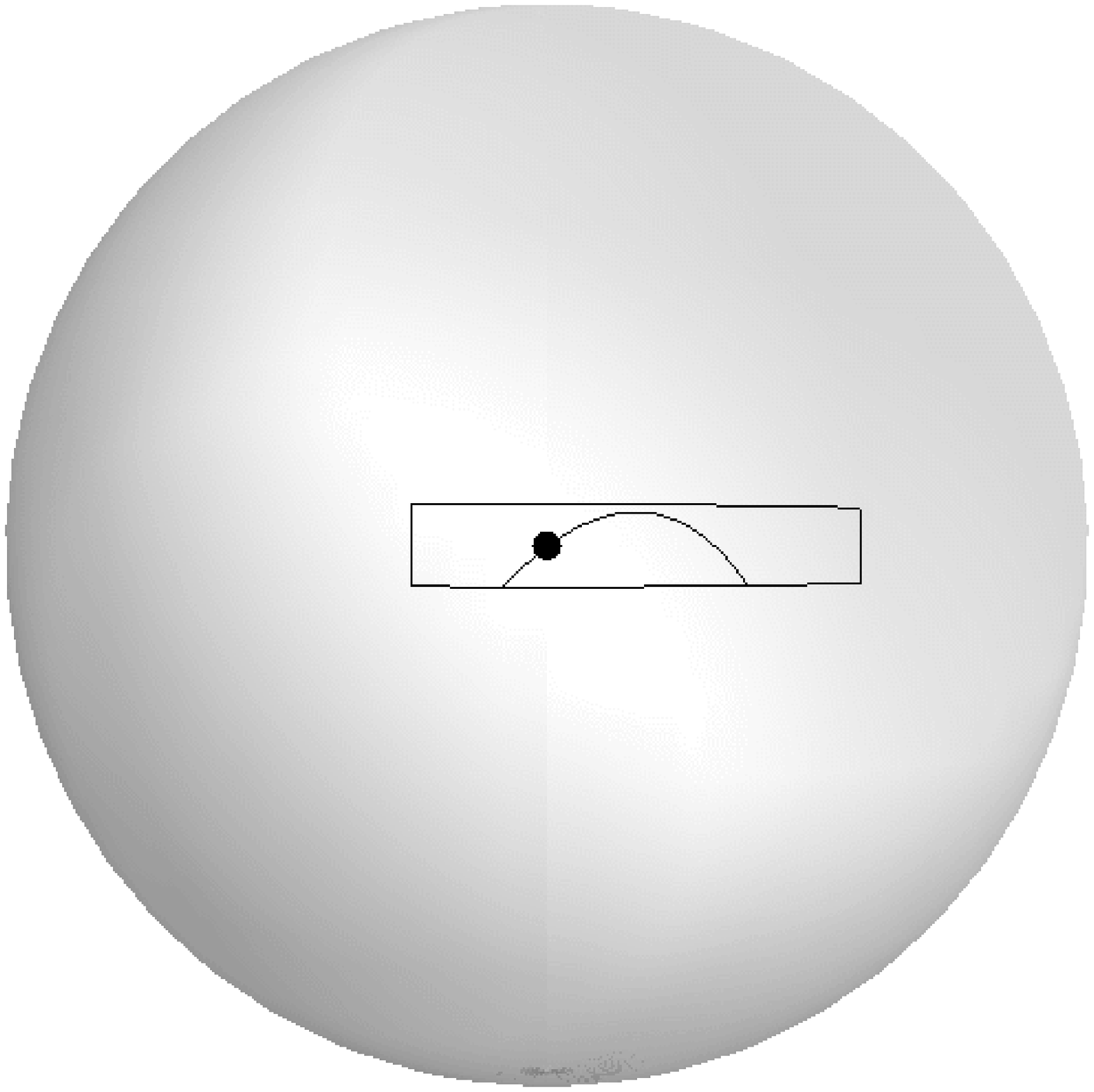}&
\hspace{-1cm}
\includegraphics[height=6cm]{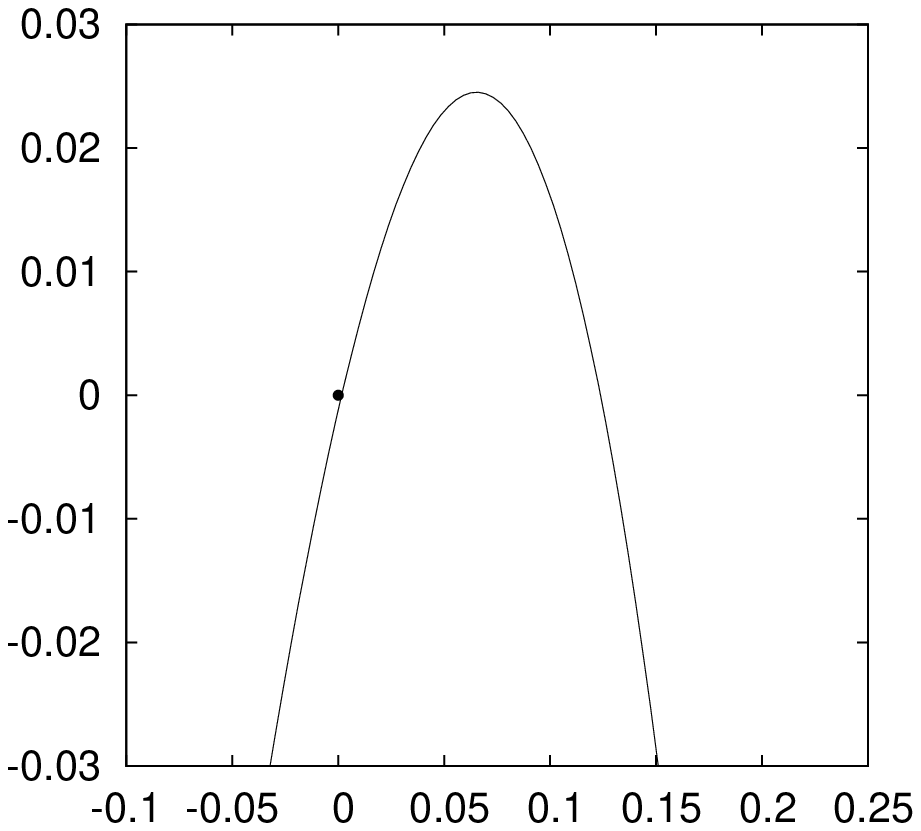}
\end{tabular}
\caption{The state vectors and $\balpha$ range over a unit sphere in $\mathbb{R}^3$.  These are conveniently represented as points on the plane under stereographic projection.  The point shown lies at the south pole of the projection.  The box on the sphere is the boundary of the plot on the right, and the curve shows the best fit line found.}
\label{fig:SP}
\end{center}
\end{figure}

Before we give any results we will take a moment to discuss their presentation.  The state vectors and $\balpha$ are in $\mathbb{R}^3$ and of unit length so take values on the surface of a unit sphere.  We will represent their positions by stereographically projecting onto $\mathbb{R}^2$.  It turns out that $\balpha$ does not need to rotate very far from $\mu=m_t$ to $\mu=m_e$ so most of the action happens in a small area on the sphere.  We have chosen the south pole of the projection to be at the position of $\mathbf{v}_\tau$.  This means that there will not be much distortion introduced by the stereographic projection; geodesics in this region on the sphere will be almost straight lines on the plane.  Figure \ref{fig:SP} shows the unit sphere with the region we will be interested in enclosed in a box.  The curve within the box shows the best fit line we find and the point shows the south pole of the projection, $\mathbf{v}_{\tau}$.  The projection itself is shown on the right.  The metric on the sphere is given by

\begin{equation}
ds^2=\frac{4}{(1+u^2+v^2)^2}(du^2+dv^2)
\end{equation}
for coordinates on the plane $u$ and $v$.  Over the boundary box in
Figure \ref{fig:SP} the metric ranges from $4(du^2+dv^2)$ to $3.76(du^2+dv^2)$; there is a maximum distortion of a length in the stereographic projection of $3\%$.

Though we assume that the trajectory of $\balpha$ is universal, there is no physical constraint on the relation between the quark and lepton sectors.  We thus have the freedom to match these two sectors to give the smoothest trajectory.  As mentioned previously we also have a freedom in choosing the $UD$ matrix.  To find the best matching with experimental data we ranged over, and then applied simplex optimization at good regions in, the parameter space of quark-lepton sector matching matrices and $UD$ matrices.  For each point in parameter space tested we found the positions of $\balpha_{t,\, c,\, b,\, s,\, \tau,\, \mu}$ and projected them onto the plane.  We then fit a cubic line to these points on the plane using a non-linear least squares algorithm.  Previous work \cite{cevidsm} has shown that the cumulative arc length between $\balpha_x$ can be modelled by an exponential function at high scales.  Accordingly, we then fit an exponential function to the cumulative arc length, excluding the strange quark and the electron.  The strange quark was excluded since the interpretation of its intermediate mass is somewhat uncertain in this scheme.  The electron was excluded firstly since later we will only be interested in the relationship at higher scales and secondly since we do not have good restrictions on the cumulative arc length, as $\balpha_e$ can only be constrained to lie on a line.  Finally we use the Jarlskog invariant to fix the value of $\theta$ and so determine the magnitudes of the elements of the CKM matrix.

\begin{figure}
\begin{center}
\begin{tabular}{cc}
\multicolumn{2}{c}{\hspace{-2cm} \includegraphics[width=13cm]{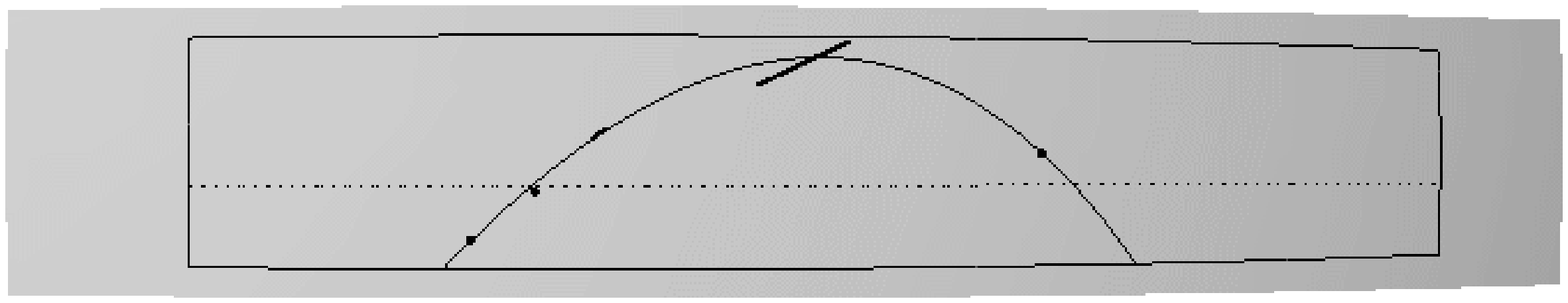} \vspace{1cm}}\\
\hspace{-1.5cm}
\includegraphics[width=8cm]{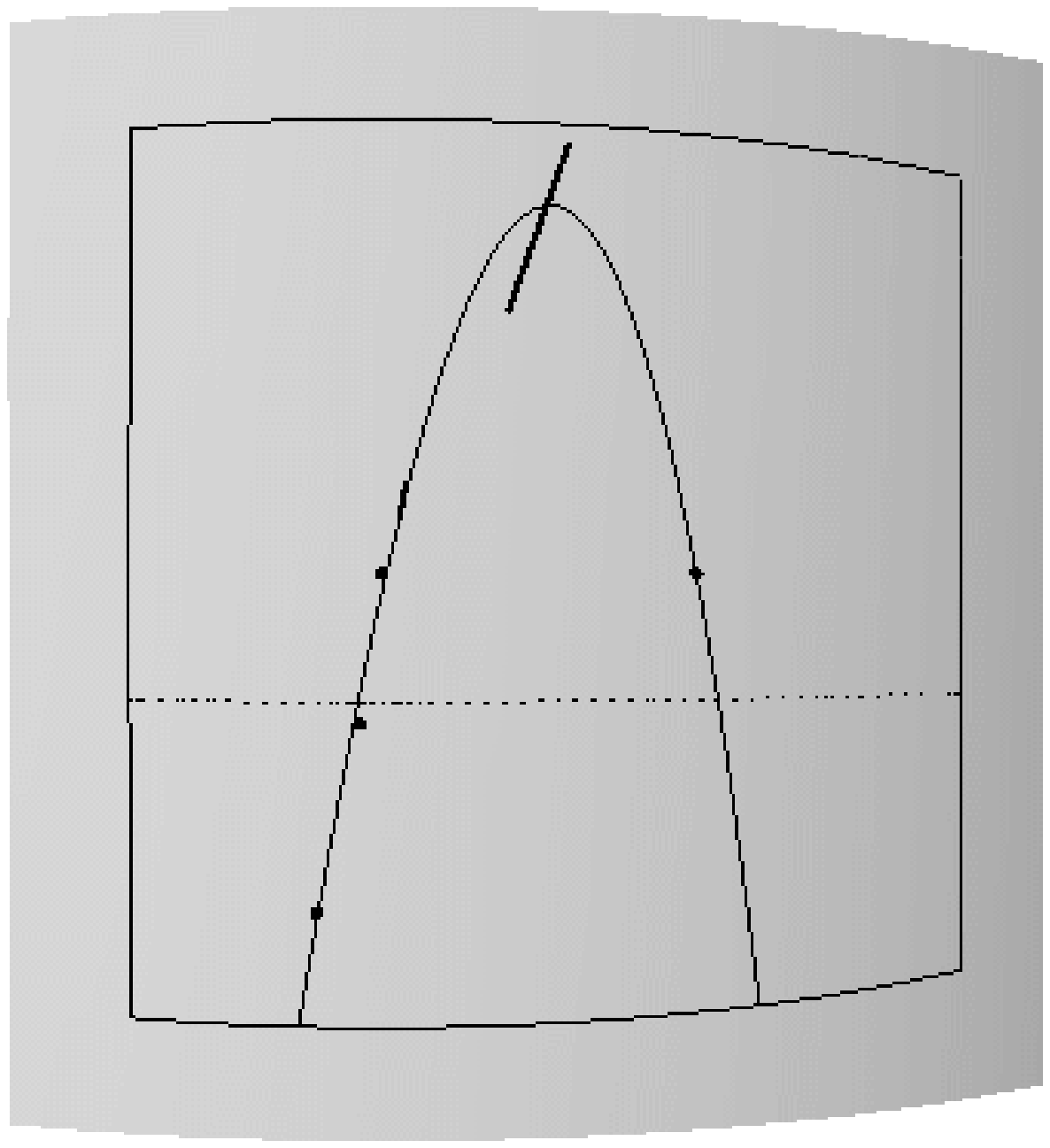}&
\hspace{-2cm}
\begin{overpic}[width=11cm]{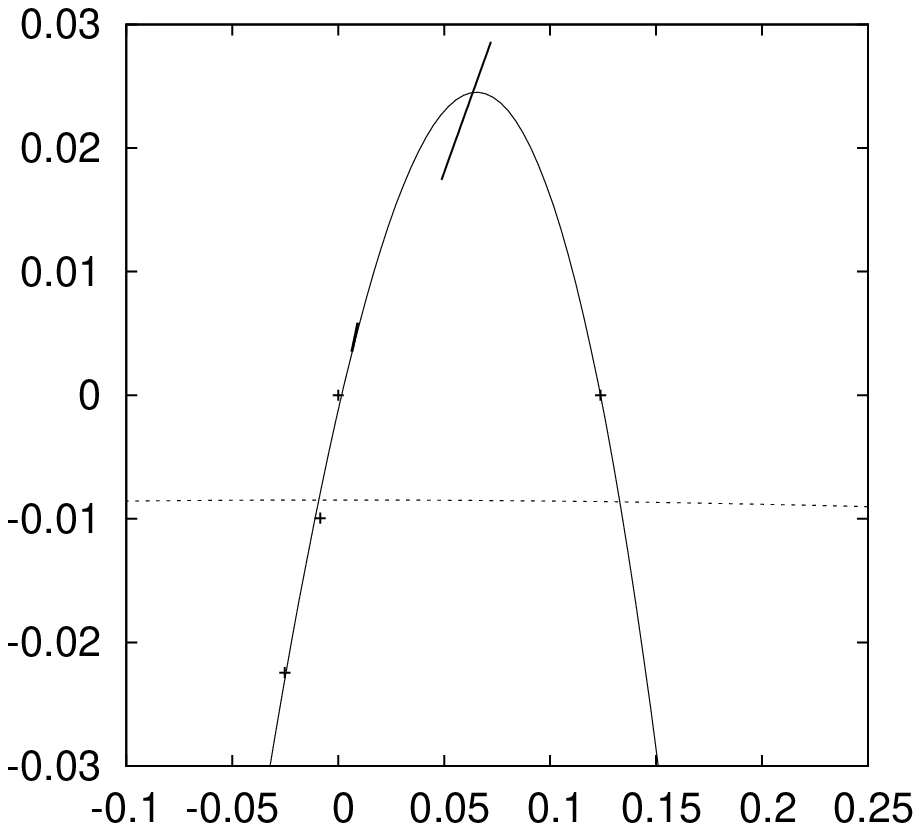}
\put(32,13){$\balpha_t$}
\put(42,24){$\balpha_b$}
\put(36,36){$\balpha_\tau$}
\put(38,41){$\balpha_c$}
\put(52,53){$\balpha_s$}
\put(65,36){$\balpha_\mu$}
\put(86,27){$\balpha_e$}
\end{overpic}
\end{tabular}
\caption{The positions of $\balpha$ at various scales determined partly by experimental constraints and partly by our choices as described in the text.  The position of $\balpha_e$ is constrained to lie somewhere on the dotted line.  The cubic best fit line is shown by the solid line.  The top shows these on the sphere, left shows them on an ellipsoid with the axes stretched to match those of the stereographic projection while the right shows the stereographic projection with the positions of $\balpha_x$ indicated.}
\label{fig:Best_Fit}
\end{center}
\end{figure}

Figure \ref{fig:Best_Fit} shows the cubic best fit line along with the positions of $\balpha_x$ found in \cite{btfit}.  The experimental errors in the masses of the quarks lead to an uncertainty in the position of $\balpha_c$ and $\balpha_s$.  The 1 $\sigma$ errors in the quark masses restrict $\balpha_c$ and $\balpha_s$ to lie on the lines shown.  The cubic best fit line is

\begin{equation}
0.75\, u-4.83\, u^2-9.19\, u^3.
\end{equation}

The cumulative arc length between $\balpha_x$'s was found after mapping the cubic best fit line onto the sphere through an inverse stereographic projection.

\begin{figure}
\begin{center}
\begin{overpic}[width=10cm]{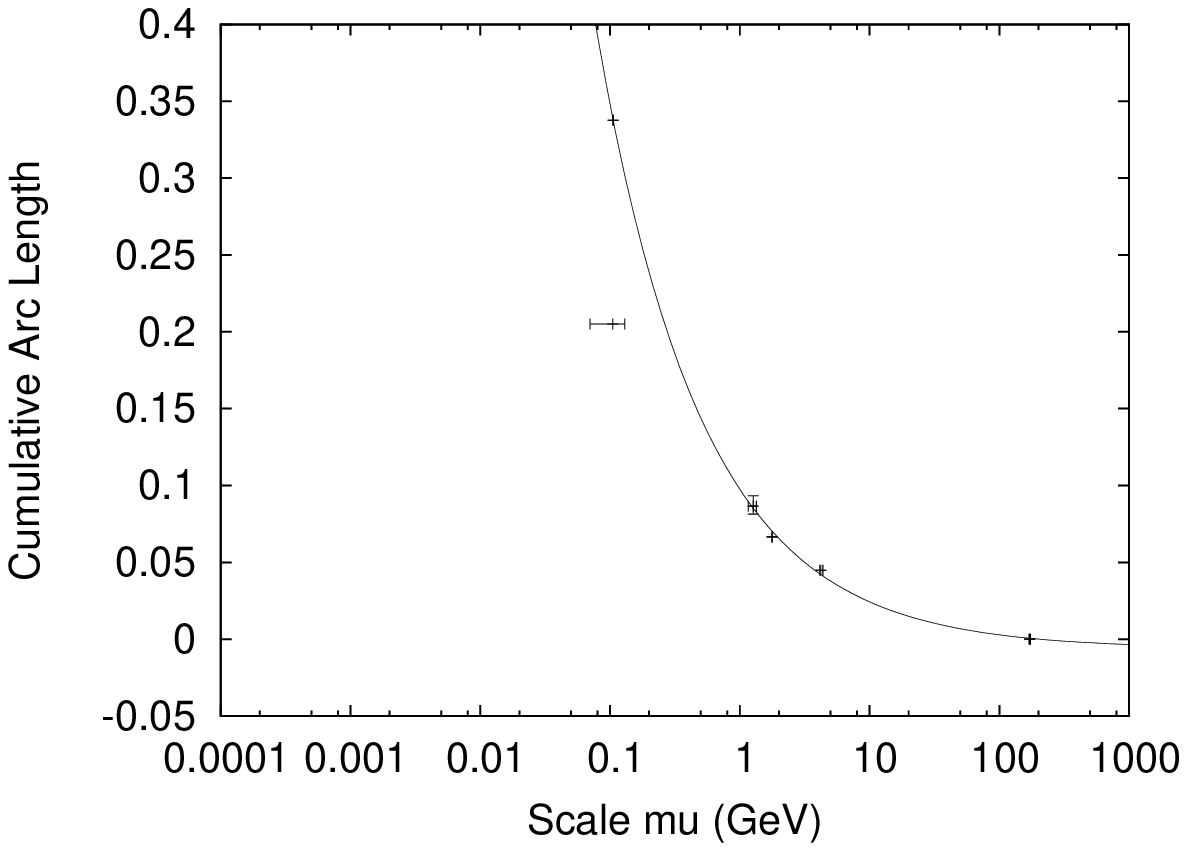}
\put(85,18){$\balpha_t$}
\put(66,19){$\balpha_b$}
\put(62,22){$\balpha_{\tau}$}
\put(59,26){$\balpha_c$}
\put(46,41){$\balpha_s$}
\put(56,58){$\balpha_{\mu}$}
\end{overpic}
\caption{The cumulative arc length along the best fit line measured from $\balpha_t$ is well approximated by an exponential curve for all but $\balpha_s$.  A great circle gives an arc length of $2 \pi$ in these units. }
\label{fig:Arc_Length_Line}
\end{center}
\end{figure}

Figure \ref{fig:Arc_Length_Line} shows that the arc lengths between $\balpha_t$, $\balpha_b$, $\balpha_{\tau}$, $\balpha_c$ and $\balpha_{\mu}$ are well fitted by the exponential curve
\begin{equation}
0.104\exp\left(-1.228\log_{10}\left(\mu\right)\right)-0.0061
\end{equation}
for $\mu$ in GeV.  This is in good agreement with the results from the 
planar approximation (Figure 3)
found in \cite{cevidsm} and provides a good fit to data points down to scales of around 100 MeV.   
For the later work on Higgs decay we only need to model the behaviour at high scales.

\begin{figure}
\begin{center}
\begin{tabular}{cc}
\hspace{-2.6cm}
\begin{overpic}[width=10cm]{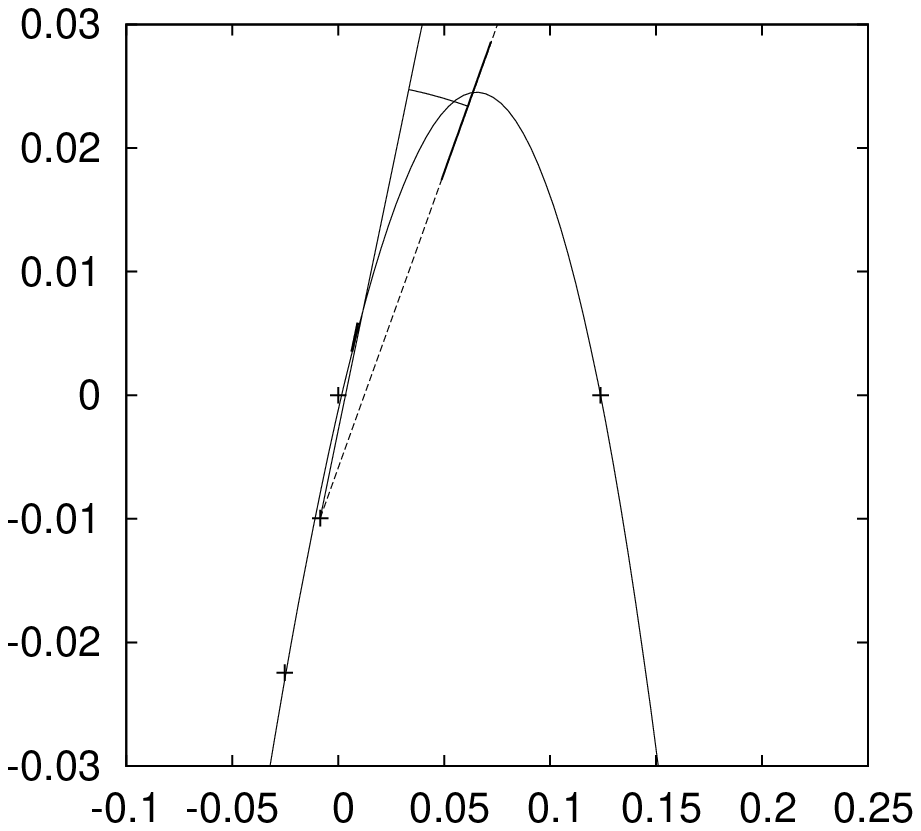}
\put(48.7,62.5){$\omega_D$}
\end{overpic}&
\hspace{-3cm}
\begin{overpic}[width=10cm]{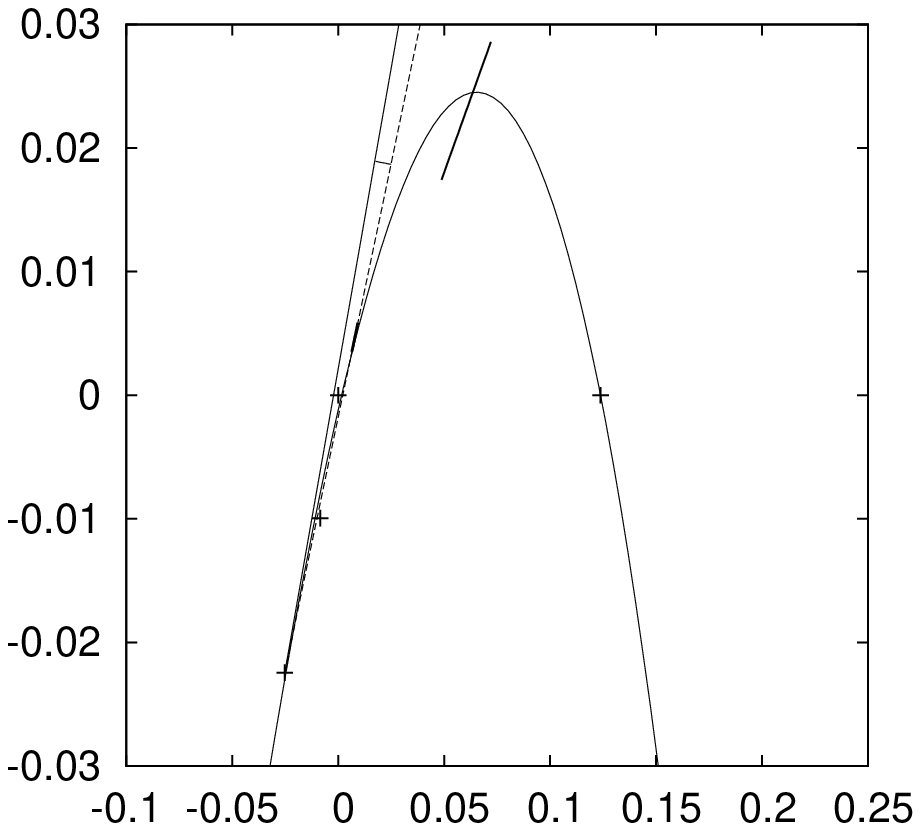}
\put(42.2,56.2){$\searrow$}
\put(37.8,59){$\omega_U$}
\end{overpic}
\end{tabular}
\caption{An illustration of the angle $\omega_D$ ($\omega_U$).  Here the state vector triad and the Darboux triad have their origins at $\balpha_{b}$ ($\balpha_{t}$).  The solid line shows $\boldsymbol{\tau}(\mu)\ ||\ \dot{\balpha}(\mu)$ and the dashed line shows $\mathbf{v}_{s}$  ($\mathbf{v}_{c}$).  It should be remembered that the axes are not of equal scale and that this is a stereographic projection of the vectors so the angles cannot be directly read off.}
\label{fig:Omega}
\end{center}
\end{figure}
For this trajectory we find $\omega_U = 0.09$ radians and 
$\omega_D = 0.25$ radians (Figure 7).  
Fitting a Jarlskog invariant of $J=3.05\times10^{-5}$ gives a strong CP angle of $1.45$ radians.  These results are in line with estimates in \cite{atof2cps}.  The absolute values of the CKM matrix obtained are:
\begin{equation}
\left(\begin{array}{ccc}
0.97430&0.2252&0.00357\\
0.2251&0.97345&0.0415\\
0.00879&0.0407&0.999134\\
\end{array}\right),
\end{equation}
which can be compared with the experimental values \cite{databook}:
\begin{equation}
\left( \begin{array}{ccc}
0.97419\pm0.00022&0.2257\pm0.0010&0.00359\pm0.00016\\
0.2256\pm0.0010&0.97334\pm0.00023&0.0415^{+0.0010}_{-0.0011}\\
0.00874^{+0.00026}_{-0.00037}&0.0407\pm0.0010&0.999133^{+0.000044}_{-0.000043}
\end{array} \right).
\end{equation}
We find the unitarity angles, defined and measured \cite{databook} as
\begin{eqnarray}
\alpha=\arg\left(-\frac{V_{td}V_{tb}^*}{V_{ud}V_{ub}^*}\right)&=&(88^{+6}_{-5})^\circ,\\
\beta=\arg\left(-\frac{V_{cd}V_{cb}^*}{V_{td}V_{tb}^*}\right)&=&\frac{1}{2}\sin^{-1}(0.681\pm0.025),\\
\gamma=\arg\left(-\frac{V_{ud}V_{ub}^*}{V_{cd}V_{cb}^*}\right)&=&(77^{+30}_{-32})^\circ,
\end{eqnarray}
to be $\alpha=88^\circ$, $\sin(2\beta)=0.691$ and $\gamma=70^\circ$.

From the above analysis \cite{btfit}, one concludes that the recent 
data cited
can all be accommodated, to within the impressive accuracy achieved
in experiment and in detail, by a smooth trajectory for $\balpha$ 
as required by R2M2.  It can be noted there that the CP-violating 
phase, introduced in the manner peculiar to the R2M2 framework as 
detailed in the preceding subsection, plays a role in achieving 
this result.  Further, as a by-product, the fitted trajectory will
be of use to phenomenological applications of the hypothesis, an 
example of which will follow in the next subsection.

\subsection{The prospect of a charmless Higgs}

In \S2.1, use is made of a special feature of R2M2, namely that of 
maintaining chiral invariance while leaving all quarks massive, to 
offer a solution to the strong CP problem which, lying as it does 
outside the original remit of R2M2, lends strong support to its 
validity.  This feature is however but a consequence of rotation.
Can rotation itself then not give direct consequences which can be
tested against experiment?  At first sight, the answer would seem 
to be simple.  The rotation of the mass matrix implies that what 
is a mass eigenstate at one scale would no longer appear as one at 
another scale but as some linear combination of mass eigenstates.  
The effects would seem thus to lead to flavour-violations and be 
easily identifiable.  Indeed some early efforts have been made 
\cite{impromat} to try to explore some of them.  These, however,
involve an assumption which R2M2 itself cannot justify and which 
under closer scrutiny cannot easily be maintained.  The rotation of 
the mass matrix, like the running of other quantities, is supposed 
to be a consequence of renormalization, presupposing therefore an 
underlying renormalizable theory.  Even though one can treat this
rotation, as one has done in this paper, as a hypothesis, without 
specifying the theory on which it is based, there is a tacit limit 
to its application.  The underlying theory, if one knows what it 
is, will give rise, presumably, to other renormalization effects 
besides the rotation of the mass matrix, and if the investigation
one is pursuing involves those other effects, they may counteract 
or otherwise modify the effects deduced from the rotating mass 
matrix alone, making them thereby erroneous.  Indeed, it was shown 
\cite{transmudsm}, using an early phenomenological model \cite{ckm}, 
that some effects on flavour violations deduced from the rotation 
of the mass matrix alone would be cancelled automatically by an 
opposite rotation effect due to wave function renormalization.  The 
mass spectrum and mixing matrices evaluated above with R2M2 alone 
just happen to be particularly simple ``static'' single-particle 
properties which, we believe, are not affected by other effects of
renormalization.  For anything beyond, however, one has to be wary 
in making predictions with R2M2 alone, for which only a calculation 
with a full-blown theory of all renormalization effects can really 
be a fool-proof guarantee.

Defying this caution, however, let us consider in the R2M2 context 
the example of Higgs boson decay into fermion-antifermion pairs.  
This being still, in a sense, a static property, it may perhaps 
survive better an unsophisticated treatment, and being also timely, 
in view of the LHC's imminent coming of age, it may perhaps be 
worth the risk.  At least, one has seen no evidence yet of the 
complications that one has seen before in \cite{transmudsm}.

To investigate Higgs decay into fermion-antifermion pairs, we will
need the Yukawa coupling.  One possibility which will give the 
required factorizable mass matrix (\ref{mfact}) is the following:
\begin{equation}
{\cal A}_{YK} = \rho_T  \bar{\psi} \balpha \phi_W  \balpha^{\dagger}   \psi  .
\label{Yukawa}
\end{equation}
Expanding $\phi_W$ about its minimum value $\zeta_W$, thus: $\phi 
= \zeta_W + H$, we obtain to zeroth order the fermion mass matrix 
as in (\ref{mfact}) with $m_T = \rho_T \zeta_W$, and to first order 
the coupling matrix of the Higgs boson to the fermions as:
\begin{equation}
\Gamma = \rho_T  \balpha \balpha^{\dagger}.
\label{Hcoup}
\end{equation}

Superficially, this result looks familiar, namely that the fermion
mass and Higgs coupling are proportional to each other.  However, 
as usually meant, the proportionality is between the mass and the 
Higgs coupling of each of the fermions individually, namely that 
$m_i = \zeta_W y_i$, with $i$ denoting the individual fermion state, 
but here it is a proportionality between matrices: $m = \zeta_W 
\Gamma$.  And both these matrices rotate.  Having seen above that 
rotation of the mass matrix $m$ alone already leads to intriguing 
consequences, we shall not be surprised that here too rotation of 
the Higgs coupling matrix $\Gamma$, which can be thought of as the
``Higgs state tensor'', will give new interesting 
results. 

Let us concentrate first on the mode into $c \bar{c}$ pair which 
gives the most dramatic effect.  As usual, for the decay amplitude, 
one would take the element of the matrix $\Gamma$
in (\ref{Hcoup}) between $c$ state vectors, i.e.
\begin{equation}
A(H \rightarrow c \bar{c}) = 
   \rho_U |\langle {\bf c}|\balpha \rangle|^2.
\label{Hdecayamp}
\end{equation}  
However, since $\balpha$ in (\ref{Hdecayamp}) now depends on $\mu$, we 
have to specify at which scale to evaluate it.  The usual convention 
is that, for Higgs decay, quantities which depend on scale should 
be evaluated at the scale of the Higgs mass, i.e.\ $\mu = M_H$, 
hence:
\begin{equation}
A(H \rightarrow c \bar{c}) = 
   \rho_U |\langle {\bf c}|\balpha(\mu = M_H \rangle|^2.
\label{Hdecayampp}
\end{equation}
The Higgs mass is, of course, still unknown, but is limited by recent
experiment to be above 115 GeV.  In other words, it is getting to 
be near the scale of the top mass $m_t \sim 170$ GeV.  In any case, 
according to Figures \ref{planarplot} and \ref{fig:Arc_Length_Line}, 
for example, $\balpha$ is fast 
approaching its asymptotic value at $\mu = \infty$ and not changing much
there already, so that effectively we have $\balpha(\mu = M_H) \sim 
\balpha(\mu = m_t)$.  Recall now, however, that $\balpha(\mu = m_t)$ 
is by definition ${\bf t}$, the state vector for $t$, and orthogonal 
to ${\bf c}$, thus giving a near zero value for (\ref{Hdecayampp}), 
the decay amplitude.  One concludes therefore that the decay width
for the mode $H \rightarrow c \bar{c}$ evaluated with R2M2 in this 
way will be much suppressed compared with the value expected in the 
standard model $\sim \rho_U^2 m_c^2$ which would be the value obtained 
from (\ref{Hdecayamp}) if the scale-dependent $\balpha$ were evaluated 
at $\mu = m_c$, not at $\mu = M_H$ as one thought it should be.  And
quite clearly, the suppression is a direct consequence of rotation.

Similar arguments obviously apply to the other fermions of the second 
heaviest generation $s$ and $\mu$, although the suppression will be 
less drastic in both these cases, since ${\bf b} = \balpha(\mu = m_b)$ 
and $\btau = \balpha(\mu = m_\tau)$ are neither of them as close 
to $\balpha(\mu = M_H)$ as ${\bf t} = \balpha(\mu = m_t)$ is.  On 
the other hand, we notice that the decay into the heaviest generation 
fermions of any type are not much affected by rotation, these being 
dependent on the cosines of the rotation angles involved, not on the 
sines of the angles as the second generation fermions are.  This means 
that the branching ratios into the second heaviest to the heaviest
generations will all be much suppressed.

It is not difficult to give some actual estimates for the suppression 
factors since from fitting the mass and mixing data before, we have 
already a fair idea of how $\balpha$ behaves as a function of $\mu$.  
For example, reading from Figure \ref{planarplot} 
or \ref{fig:Arc_Length_Line} the angle corresponding 
to $\balpha(\mu = M_H)$, which is straightforward (apart from a minor
technical point explained in \cite{anHdecay} concerning a possible 
shift in calibration of scales which makes, however, little difference 
to the conclusions) one has for $M_H = 150\ {\rm GeV}$:
\begin{equation}
\frac{\Gamma(H \rightarrow c \bar{c})}{\Gamma(H \rightarrow b \bar{b})}
   \sim 4.3 \times 10^{-7}.
\label{csupp}
\end{equation}
which is nearly 5 orders of magnitude less than the standard model 
expectation of $\sim m_c^2/m_b^2 \sim 0.09$.  Similar estimates for 
the corresponding branching ratios for $s$ and $\mu$ are, again for 
$M_H = 150\ {\rm GeV}$, respectively $1.8 \times 10^{-6}$ and $2.9 
\times 10^{-6}$ as compared to standard model expectations of about 
$6 \times 10^{-4}$ for both.

\begin{figure}
\begin{center}
\begin{tabular}{cc}\hspace{-2.5cm}
\begin{overpic}[width=9cm]{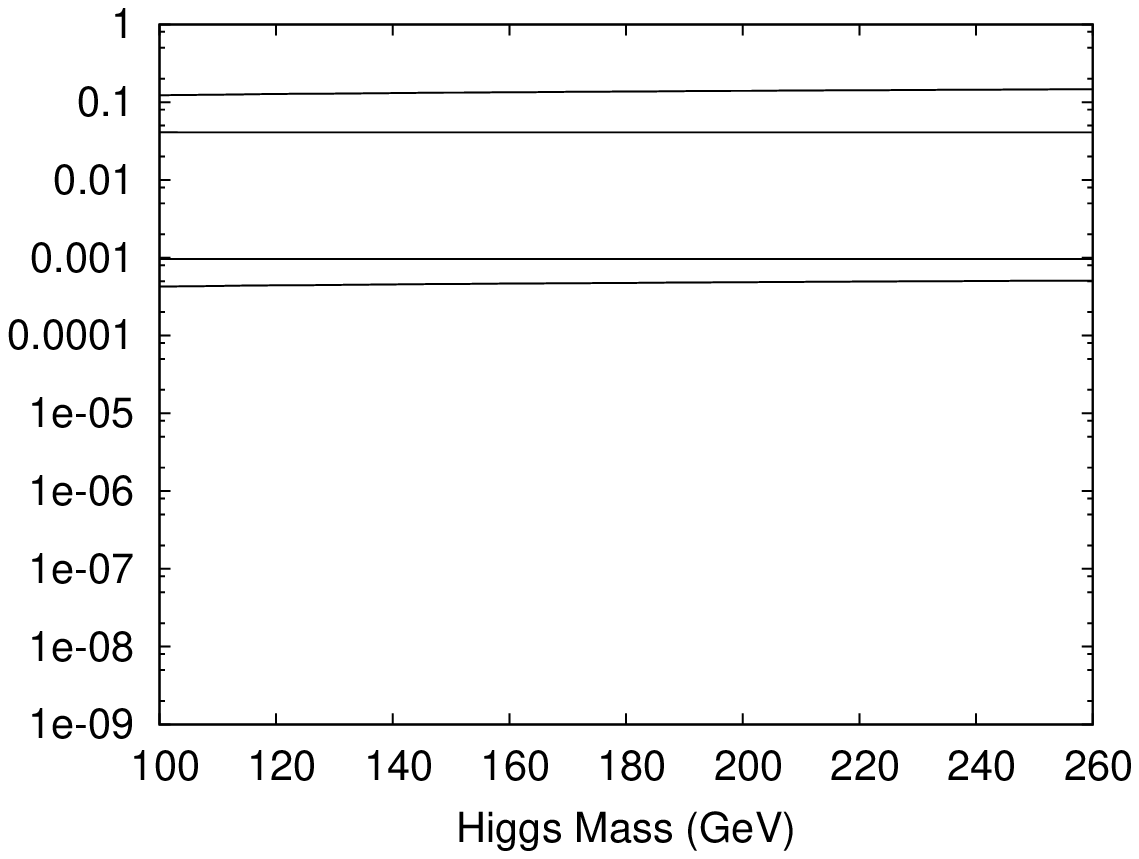}
\put(68,62){$\tau^-\tau^+$}
\put(68,53.5){$c\bar{c}$}
\put(78,49){$s\bar{s}$}
\put(78,41){$\mu^-\mu^+$}
\end{overpic}
&\hspace{-1cm}
\begin{overpic}[width=9cm]{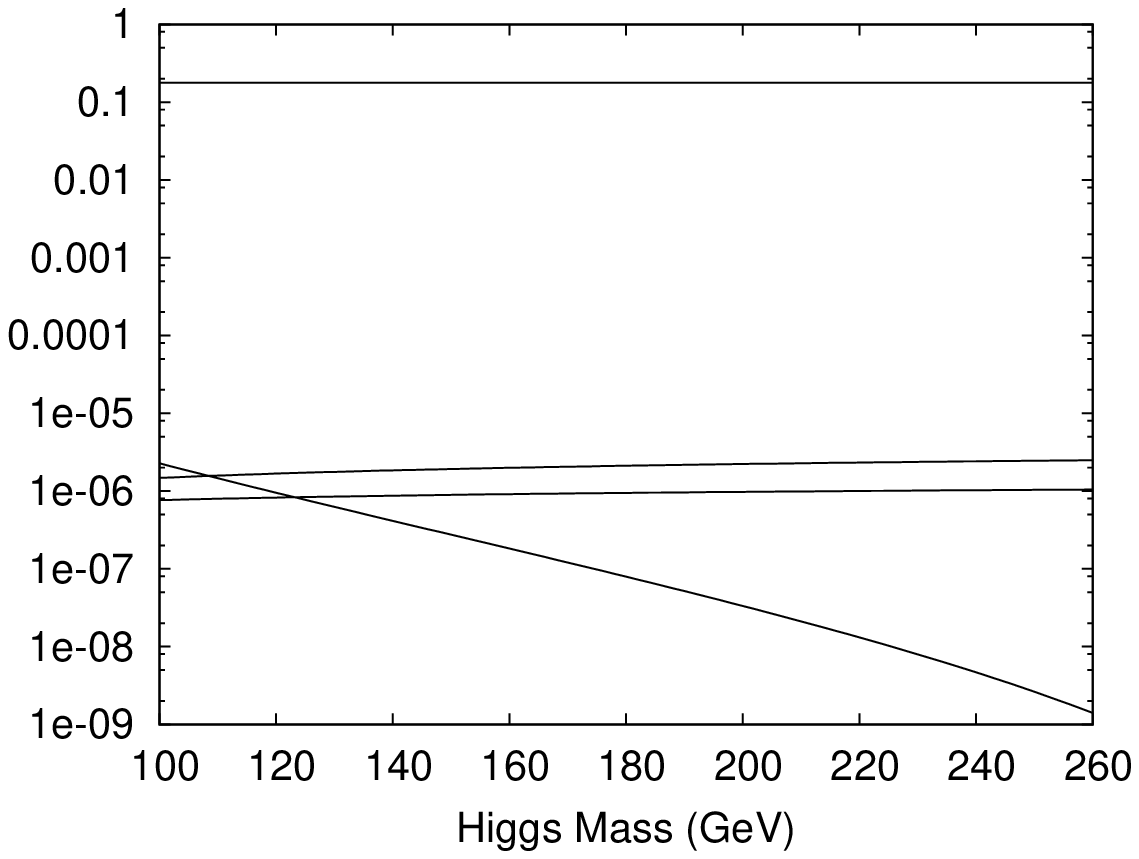}
\put(75,57){$\tau^-\tau^+$}
\put(50,18){$c\bar{c}$}
\put(75,33){$s\bar{s}$}
\put(75,24){$\mu^-\mu^+$}
\end{overpic}
\end{tabular}
\caption{$\Gamma(H\rightarrow x\bar{x})/\Gamma(H\rightarrow b\bar{b})$ for various final state particles as predicted by the standard model (left) and the rotating mass matrix hypothesis (right).}
\label{fig:BRSM}
\end{center}
\end{figure}

Making use of the more detailed 3-G fit to mass and mixing data
described in \S2.2, one will be able to derive the branching 
ratios of anomalously suppressed modes involving the lightest
generations too, although not as transparently.  Some such branching
ratios are shown in Figure \ref{fig:BRSM} on the right, with comparison
with standard model results on the left \cite{Spira:2009}.
The lightest generations had branching ratios less than $10^{-9}\times\Gamma(H \rightarrow b \bar{b})$.

\begin{figure}
\begin{center}
\begin{tabular}{cc}\hspace{-2.5cm}
\begin{overpic}[width=9cm]{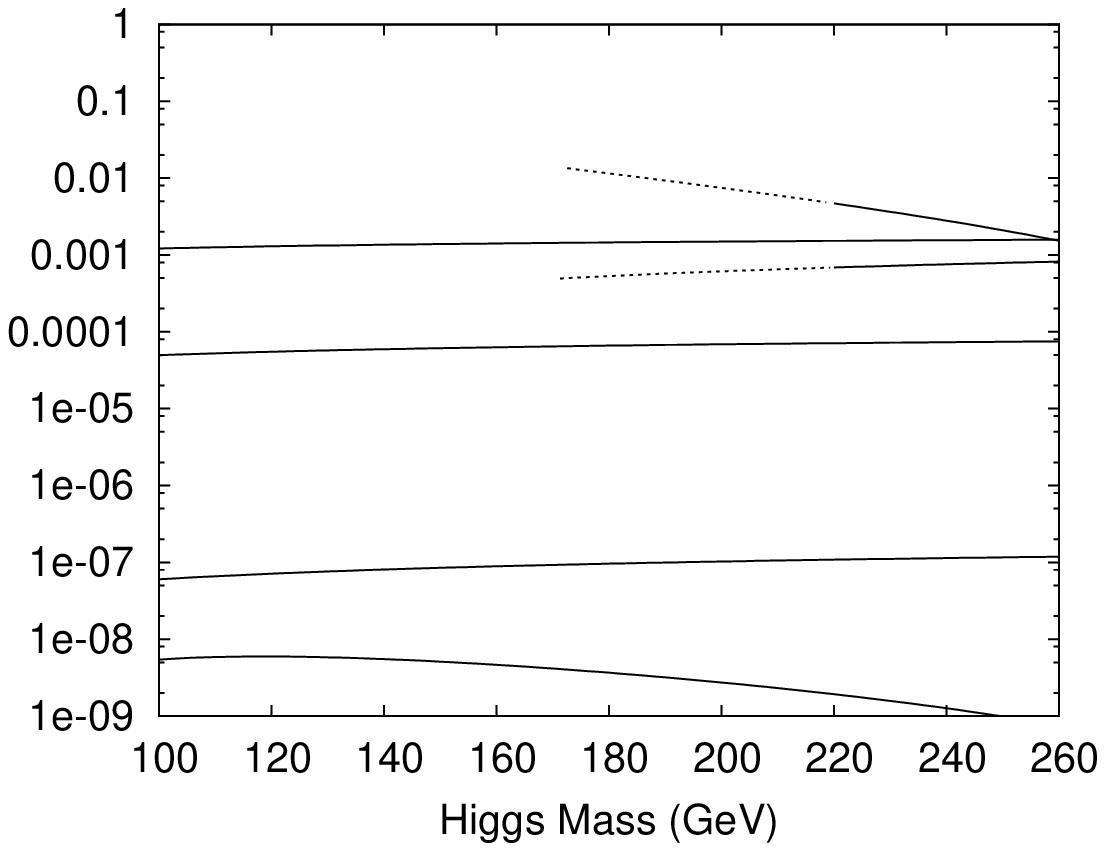}
\put(30,49){$b\bar{s}$}
\put(30,41){$b\bar{d}$}
\put(68,24){$s\bar{d}$}
\put(68,14){$c\bar{u}$}
\put(48,53){$t\bar{c}$}
\put(47,44){$t\bar{u}$}
\end{overpic}
&\hspace{-1cm}
\begin{overpic}[width=9cm]{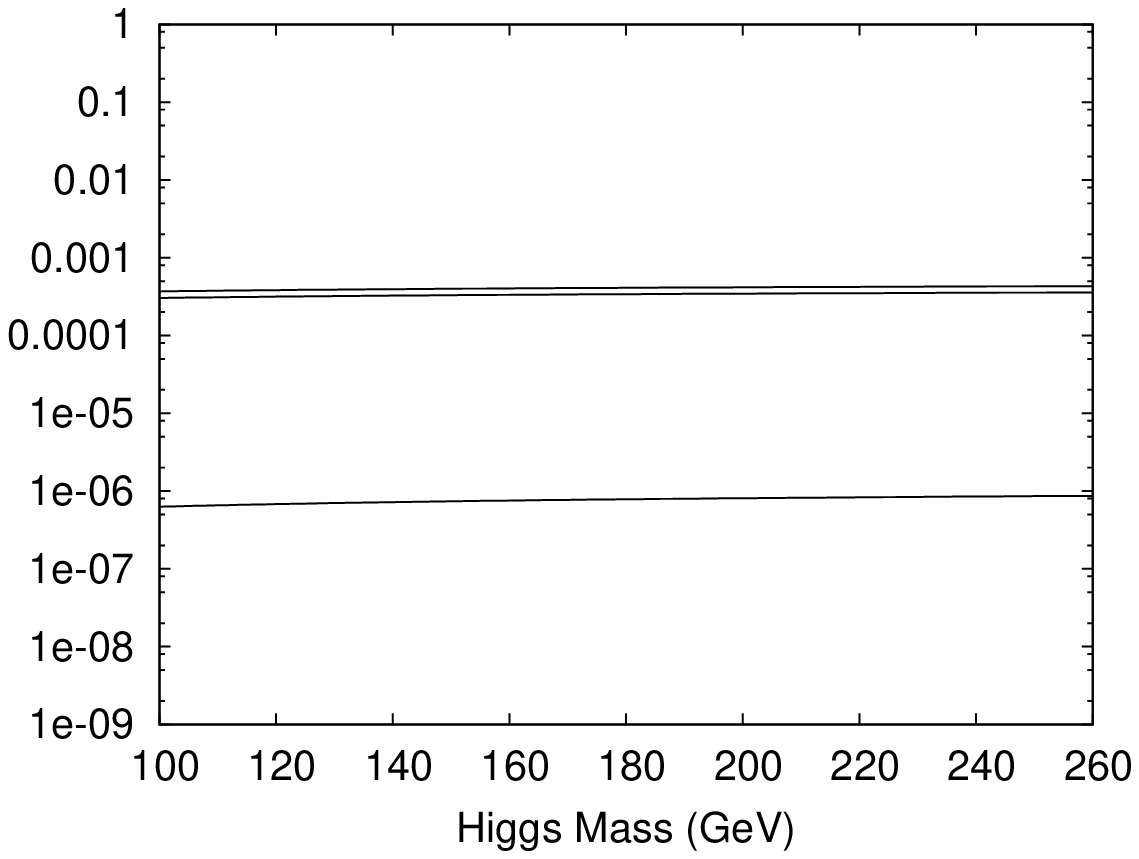}
\put(75,47){$\tau^-\mu^+$}
\put(75,40){$\tau^-e^+$}
\put(75,30){$\mu^-e^+$}
\end{overpic}
\end{tabular}
\caption{$\Gamma(H\rightarrow x\bar{y})/\Gamma(H\rightarrow b\bar{b})$ for various flavour violating decays as predicted by the rotating mass matrix hypothesis.  Note that $\Gamma(H\rightarrow y\bar{x})=\Gamma(H\rightarrow x\bar{y})$.  Below around $220$ GeV, indicated by the dotted lines, threshold effects will influence the $t\bar{c}$ and $t\bar{u}$ decay modes.}
\label{fig:BRFV}
\end{center}
\end{figure}

Apart from the modes considered above which are anomalously
suppressed, there is another category of anomalous modes with
a similar origin.  The Yukawa coupling in (\ref{Hdecayampp}) 
for a general $\mu$ has usually nondiagonal matrix elements 
between state vectors of fermions of the same type but from
different generations, leading to flavour-violating decays of
the sort: $H \rightarrow \tau \bar{\mu}$.  Branching ratios 
predicted from Figure \ref{planarplot} or the fit in \S2.2 for 
such modes are generally quite small, for example, from the
former, one has, again for $M_H=150$ GeV:
\begin{equation}
\frac{\Gamma(H \rightarrow \tau \bar{\mu})}
   {\Gamma(H \rightarrow b \bar{b})} = \frac{m_{\tau}^2}{m_b^2}
   \frac{\cos^2 \theta_{H \tau} \sin^2 \theta_{H \tau}}
   {\cos^4 \theta_{H b}} \sim 7.7 \times 10^{-4}.
\label{BRtaumu}
\end{equation}
which, being so distinctive though small, may eventually be seen.
The more general conclusions from the fit in \cite{btfit} is 
summarized in Figure \ref{fig:BRFV}.

Predictions on flavour-violation are often troublesome in that
the effect may propagate and lead to violations in other areas
already very strongly bounded by experiment.  Of most danger
in this case that one can identify is the implication of Higgs
exchange for the mass difference between the two neutral-strange 
$B$ mesons, for which one obtains from Figure \ref{planarplot}
the following:
\begin{equation}
m_{B^0_{sH}} - m_{B^0_{sL}} \sim 4.8 \times 10^{-10}\ {\rm MeV}
\label{DeltamBse}
\end{equation}
which is more than an order of magnitude below the experimentally
measured value of $117 \times 10^{-10}\ {\rm MeV}$.  Given the
difficulty of making accurate theoretical predictions for such
hadronic quantities, the Higgs contribution is unlikely to be
noticeable at present and thus causes no problem, but it may be 
something to look for in future when experimental measurement and 
theoretical interpretation both continue to improve.  As for other 
implied flavour-violations like $\tau \rightarrow \mu \mu \bar{\mu}$ 
decay, the estimates are so low as to cause no worry for a long time 
to come.  Thus, the conclusion for the moment is that the prediction
of flavour-violation in Higgs decay, such as (\ref{BRtaumu}) above,
does not lead to contradiction to experiment elsewhere.

For more details and more examples of similar results, the reader
is referred to the original research papers \cite{anHdecay,btfit}. 

These two ``predictions'', namely of the anomalous suppression of
modes like $c \bar{c}$ and of flavour-violations in modes like
$\tau \bar{\mu}$, are exciting in that they are exotic and can 
in principle be checked soon against experiment at the LHC,
although, we are told, some channels may initially be difficult
to identify.  However, they are by no means as solid as the
result in \S2.1, depending as they do on further assumptions the
validity of which one is in no position at present to assess.
Indeed, as far as these ``predictions'' are concerned, R2M2 seems
to be placed in a, perhaps fortunate but certainly undignified,
``win-win'' situation.  If these ``predictions'' were to be confirmed 
by experiment, then R2M2 may claim success since no other scheme
is known to give such predictions.  On the other hand, if they
were to be disproved by experiment, one can always lay the blame
on the additional assumptions which have been thrown in to derive 
the results and learn instead from the failure.  Only time and 
further work, however, will be able to clarify the situation.

\section{Summary and Remarks}

The R2M2 hypothesis has two components: (a) that the fermion mass
matrix is of rank one, (b) that this matrix rotates.  The first is
old, having been suggested some 30 years ago; what is relatively new 
is the second, namely that the mass matrix changes orientation in 
generation space with changing scale.  And this simple addition is 
seen immediately to lead to (i) a hierarchical mass spectrum and (ii) 
mixing matrices for fermions with the qualitative features observed
in experiment.  A closer examination confirms that the hypothesis
is fully consistent with current data, in detail and to within the 
impressively small experimental errors recently achieved.  Exceptions,
such as the masses of the light quarks $u$ and $d$, are seen not so
much as inconsistencies than as features which the hypothesis have 
so far failed to explain due to the common lack of understanding at
present of colour confinement, inhibiting thereby direct comparison 
of the hypothesis to experiment.

An unusual consequence of the R2M2 hypothesis is that chiral symmetry
can be maintained in the action while still allowing all fermions to 
have nonzero though hierarchical masses.  Apart from repercussions
probably elsewhere, it is shown in particular to offer (iii) a novel 
solution to the strong CP problem by linking it via rotation to the 
CP-violating phase in the CKM matrix, even giving (iv) a Jarlskog 
invariant in the CKM 
matrix of the order $3 \times 10^{-5}$ experimentally observed, for a 
theta-angle of order unity in the QCD action.
This
is a particularly gratifying result in that it is obtained in an
entirely
different domain of physics than that for which the R2M2 hypothesis was
originally intended, and could thus be regarded as a nontrivial check on 
the hypothesis.  The result is also of some theoretical significance 
in that it has linked together the theta-angle of topological origin 
in QCD to the KM phase in the CKM matrix belonging to the weak current, 
two concepts previously thought to be entirely separate. 

Further, R2M2 is seen to predict (v) anomalies in Higgs decay which 
should be testable at LHC in the near future, although these depend
on further assumptions.

Taken together, the results (i)---(v) seem a fair recommendation for
R2M2 to be considered as a means to an end by model builders who are
looking behind the standard model with the aim of explaining its many 
idiosyncrasies.  Any model, it seems, which can produce a rank-one
fermion mass matrix rotating with changing scale in generation space
at sufficient speed, would have a fair chance of success.  For this 
reason, although we ourselves have suggested one such model with
R2M2 as a consequence \cite{efgt,dfsm,prepsm}, in the present review 
R2M2 is cast deliberately as a phenomenological hypothesis and only
those results derivable from R2M2 are discussed which are independent
of how the rotation is generated.  This will make it more useful, we
hope, as a guide for fellow model builders starting from different
premises with completely different ideas from ours.

\end{document}